\documentclass[aps,pre,twocolumn,superscriptaddress]{revtex4}
\usepackage[dvipdfmx]{graphicx}
\usepackage{ulem}
\usepackage{color}
\usepackage{amsmath,amssymb}
\def\vector#1{\mbox{\boldmath $#1$}}

\begin{document}
\title{Anisotropy-temperature phase diagram for the two-dimensional dipolar Heisenberg model with and without magnetic field}

\author{Hisato \surname{Komatsu}}
\email[Email address: ]{komatsu.hisato@nims.go.jp}

\affiliation{Research Center for Advanced Measurement and Characterization, National Institute for Materials Science, Tsukuba, Ibaraki 305-0047, Japan}

\author{Yoshihiko Nonomura}
\email[Email address: ]{nonomura.yoshihiko@nims.go.jp}

\affiliation{International Center for Materials Nanoarchitectonics,  National Institute for Materials Science,  Tsukuba, Ibaraki 305-0044, Japan}

\author{Masamichi Nishino}
\email[Email address: ]{nishino.masamichi@nims.go.jp} 

\affiliation{Research Center for Advanced Measurement and Characterization, National Institute for Materials Science, Tsukuba, Ibaraki 305-0047, Japan}

\affiliation{Elements Strategy Initiative Center for Magnetic Materials, National Institute for Materials Science, Tsukuba, Ibaraki 305-0047, Japan}

\begin{abstract}

We investigate phase transitions in the two-dimensional dipolar Heisenberg model with uniaxial anisotropy with a specific ratio between the exchange and dipolar constants, $\delta=1$. We obtain the $\eta$--$T$ (anisotropy vs. temperature) phase diagrams for typical values of magnetic field by a Monte Carlo method with an $O(N)$ algorithm. We find that at lower fields, the $\eta$--$T$ phase diagram consists of the planar ferromagnetic (F), (perpendicular) stripe-ordered (SO), and paramagnetic (P) phases, and is characterized by the triple point. 
In the SO phase realized at larger $\eta$ and smaller $T$, the SO pattern changes depending on the field. On the other hand, we find that at higher fields, the SO phase does not exist, while the planer F phase robustly remains. 
We study the properties of the phase boundaries by employing finite-size-scaling analyses. We find that the slope of the spin-reorientation-transition line is positive with and without field, i.e., $\frac{d\eta}{dT}>0$, which implies that the planar F phase changes to the SO phase with lowering temperature. 
In the phase diagrams we observe a characteristic shape of the P--planer F phase-transition line, whose maximum point of $\eta$ is located at an intermediate temperature. This structure leads to the temperature-induced reentrant transition associated with P and planar F phases, 
which appears in successive phase transitions with lowering temperature: 
P $\rightarrow$ planar F $\rightarrow$  P $\rightarrow$ SO phase at lower fields and P $\rightarrow$ planar F $\rightarrow$  P phases at higher fields. 
\end{abstract}

\maketitle

\section{Introduction}
\label{Introduction}

Ultrathin magnetic films exhibit a variety of orderings 
due to the competition between magnetic anisotropies, short-range exchange and long-range dipolar interactions. 
They are not only of scientific interest but also of technological importance because of potential applications such as magnetic recording. 
The dipolar interaction in planar systems induces in-plane antiferromagnetic (AF) order, the exchange interaction promotes ferromagnetic (F) order, and the uniaxial anisotropy favors Ising-spin (perpendicular to the plane) configurations. 
Such a complicated situation gives rise to experimentally observed spin-reorientation (SR) transitions between in-plane and out-of-plane magnetic phases as temperature or thickness of the film changes~\cite{Pappas,Allenspach,Ramchal,Won,Qiu}. 

Due to such complexity, phase transitions in the magnetic films including SR have not been well understood. The long-range nature of the dipolar interaction is the main obstacle for theoretical and computational studies. Simulations of $N$-spin systems with short-range interactions cost $O(N)$ computational time. However, usually $O(N^{2})$ computational time is necessary for those with long-range interactions, and analyses of large systems are much more difficult. 

In spite of this difficulty, the two-dimensional (2D) dipolar Heisenberg model with uniaxial anisotropy (see Eq.~\eqref{Hamiltonian}) has been intensively studied for understanding ultrathin-film magnetism theoretically and computationally~\cite{Pescia,Moschel,Hucht,MacIsaac1,MacIsaac2,Bell,Santamaria,Rapini,Whitehead,Carubelli,Mol,Pighin2,Pighin,Mol2}, as well as the 2D dipolar Ising ferromagnet~\cite{Booth,MacIsaac-Ising,Toloza,Rastelli06,Cannas,Pighin-Ising,Rastelli,Vindigni,Rizzi,Fonseca,Ruger,Horowitz,Bab,Komatsu}. 
Especially, the $\eta$--$T$ (uniaxial anisotropy vs. temperature) phase diagram of the model~\eqref{Hamiltonian} at zero field ($H=0$) has been focused on, while the phase diagrams for finite fields are almost unexplored.

\begin{figure}[tbh]
\begin{center}
\includegraphics[width = 6.0cm]{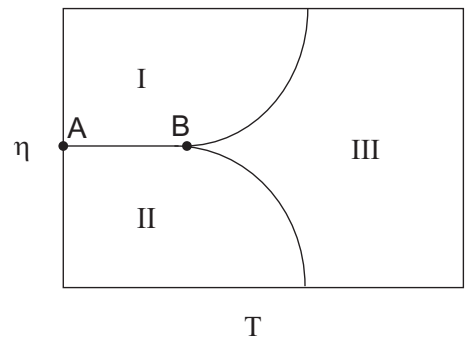}
\end{center}
\caption{Schematic $\eta$--$T$ phase diagram at $H=0$ for the model~\eqref{Hamiltonian}. Phases I and II are ordered ones whose properties depend on $\delta$, and phase III is the paramagnetic one.}
\label{rough_diag}
\end{figure}
 
The schematic $\eta$--$T$ phase diagram at $H=0$ is given in Fig.~\ref{rough_diag}. It is characterized by the triple point B. 
Phases I and II are ordered ones depending on the value of the ratio $\delta$ between the exchange and dipolar constants (see Eq.~\eqref{Hamiltonian}), and phase III is the paramagnetic (P) one. 
For $\delta=0$, i.e., the case of no exchange interaction,  phases I, II, and III are the perpendicular AF, planar AF, and P phases, respectively~\cite{MacIsaac1,Carubelli}. 
On the other hand, for $\delta \gg 1$, strong limit of the exchange interaction, phases I, II, and III are the perpendicular F, planar F, and P phases, respectively~\cite{Moschel,Leib}. 

To characterize the SR transition line (AB) between the ordered phases (phases I and II), the sign of the slope $\frac{d\eta}{dT}$ is crucial. 
For $\frac{d\eta}{dT}>0$, the SR transition occurs from phase II to phase I with lowering temperature, while for $\frac{d\eta}{dT}<0$, from phase I to phase II. However, the sign often becomes a controversial topic. 
 For $\delta \gg 1$, $\frac{d\eta}{dT}>0$ was reported~\cite{Moschel,Pescia} but for $\delta=0$, both $\frac{d\eta}{dT}>0$~\cite{Carubelli} and $\frac{d\eta}{dT}<0$~\cite{MacIsaac1} were pointed out.

For intermediate values of $\delta$ at $H=0$, perpendicular stripe-ordered (SO) phases appear as phase I. 
The phase diagrams in this case have been studied for limited values of $\delta$. 
For  $\delta=3$, the SO, planar F, and P (or tetragonal) phases appear in the phase diagram, in which the $\langle 4 \rangle$ stripe order is realized in the SO phase. 
Here we classify SO phases by using the notation introduced in Ref.~\cite{Pighin}: $\langle h_1^{n_1} h_2^{n_2} \cdots h_m^{n_m}  \rangle$, in which $h_i$ is the width of a stripe and $n_i$ is the number of consecutive stripes with the same width $h_i$. The width is measured in units of the lattice constant. 
There also exists controversy about the slope of the SR transition line for $\delta=3$: $\frac{d\eta}{dT}<0$ by MacIsaac {\it et al}~\cite{MacIsaac2} and $\frac{d\eta}{dT}>0$ by Carubelli {\it et al}~\cite{Carubelli}. 
The nature of the phase transition between the P and SO phases is also a delicate issue. Indeed, the former authors suggested a second-order transition but the latter authors a first-order transition.

In the present work, we study the $\eta$--$T$ phase diagram for $\delta=1$ 
with and without field. To reduce the difficulty of the simulation, we adopt an $O(N)$ Monte Carlo (MC) algorithm for long-range interaction spin models, called the stochastic cutoff (SCO) method~\cite{Sasaki},  with a modification efficient for the models with the uniaxial anisotropy, which was adopted in our previous study on the dipolar Ising ferromagnet~\cite{Komatsu}.

Finite-size scaling methods are powerful tools to investigate details of phase transitions.  So far those approaches have been applied in limited number of studies~\cite{Mol,Mol2}. 
However, to settle the above-mentioned controversies, such systematic approaches are more and more important. Here we investigate the phase diagram with a finite-size-scaling approach. 

We find the $\langle 1 \rangle$, $\langle 21^3 \rangle$, and $\langle 21  \rangle$ SO phases at $H=0$, 1.3, and 2, respectively, for relatively large $\eta$. These three SO phases are consistently realized in the $H$--$T$ phase diagram for $\eta \rightarrow \infty$, i.e., the dipolar Ising ferromagnet~\cite{Komatsu}. 
For intermediate $\eta$, the SR transition between the planar F and SO phases is observed except for high fields.  
We find that $\frac{d\eta}{dT}>0$ holds for the SR transition line for $\delta=1$ with and without field. 

Frustrated interactions often causes complex phase transitions. 
The temperature-induced reentrant transition observed in spin-glass systems is a typical example~\cite{Maletta,Aeppli,Childress}. Here we use the term ``reentrant transition" in a narrow sense, namely the phase changes as $A$ $\rightarrow$ $B$ $\rightarrow$ $A$. 
In the reentrant transition in spin glasses, phase $A$ is a disordered one. 
The main origin of the reentrant transition is the entropy effect but the detailed mechanisms are generally complicated. 
The reentrant transition associated with P and F phases was shown in successive transitions in a frustrated Ising model such as P $\rightarrow$ F $\rightarrow$  P phase, or P $\rightarrow$ F $\rightarrow$ P $\rightarrow$ AF phase with decreasing temperature~\cite{Kitatani}.

In dipolar systems, the competition between the anisotropy, short-range, and 
long-range interactions may cause new types of reentrant transitions. 
Recently we found a field-induced reentrant transition: uniform (paramagnetic) $\rightarrow$ $\langle 21 \rangle$ SO $\rightarrow$ uniform (paramagnetic) phase in the 2D dipolar Ising ferromagnet with $\delta=1$~\cite{Komatsu}, which corresponds to the model~\eqref{Hamiltonian} with $\eta \rightarrow \infty$. 
In the present work we find a temperature-induced reentrant transition associated with P and planer F phases for specific values of $\eta$. 
With decreasing temperature, the following changes are observed: P $\rightarrow$ planar F $\rightarrow$ P $\rightarrow$ SO phase at relatively low fields and P $\rightarrow$ planar F $\rightarrow$ P phase at relatively high fields. 

The rest of the paper is organized as follows. 
The model and methods are briefly explained in Sec.~\ref{model}. 
The phase diagrams are overviewed in Sec.~\ref{overview}. 
The phase transition between the P and planar F phases is discussed in Sec.~\ref{PF_and _D}. 
The characteristics of the SR transition between the planar F and SO phases are presented in Sec.~\ref{SR-transition}. 
The phase transition between the P and SO phases is argued in Sec.~\ref{Stripe-D}. The temperature-induced reentrant transition is discussed in Sec.~\ref{Reentrant}. 
Section~\ref{summary} is devoted to the summary.

\section{Model and methods}
\label{model}

The 2D dipolar Heisenberg model with the uniaxial anisotropy is given by 
\begin{eqnarray}
{\cal H} & = & - \delta \sum _{\left< i,j \right> } \vector{S} _i \cdot \vector{S} _j + \sum _{i < j  } \left\{ \frac{\vector{S} _i \cdot \vector{S} _j }{r_{ij} ^3} - \frac{ 3(\vector{S} _i \cdot \vector{r}_{ij} ) (\vector{S} _j \cdot \vector{r}_{ij}) }{r_{ij} ^5} \right\} \nonumber \\
& & - \eta \sum _i \left( S _i ^z \right) ^2 - H \sum _i S _i ^z ,
\label{Hamiltonian}
\end{eqnarray}
where $\delta (>0)$ is the ratio between the exchange and dipolar constants, $\eta$ is the single-ion anisotropy constant, and $H$ is the magnetic field. We consider $N=L \times L$ sites. 
The first sum $\left< i,j \right>$ runs over all nearest-neighbor pairs of spins and the second one over all pairs of spins on a square lattice. 
The spin variables $\vector{S} _i = (S_i ^x , S_i ^y , S_i ^z )$ are normalized as $| \vector{S} _i | =1$. 
The distance between sites $i$ and $j$, $r_{ij}$, is measured in units of the lattice constant. 
This model corresponds to the 2D dipolar Ising ferromagnet for $\eta \rightarrow \infty$.  

To treat large systems and exclude the effect of edges, we use a replica method which tiles replicas of the original system with periodic boundary conditions~\cite{Ruger,Horowitz,Komatsu}. We tile 2001 $\times$ 2001 replicas. 

To study the planar F phase, we define the transverse magnetization:\begin{equation}
M_{xy} = \frac{1}{N} \left \langle \sqrt{ \left( \sum _i S_i ^x \right) ^2 + \left( \sum _i S_i ^y \right) ^2 } \right \rangle 
\label{Mplanar}
\end{equation}
and the longitudinal magnetization: 
\begin{equation}
M_{z} = \frac{1}{N}  \left \langle  \sum _i S_i ^z  \right \rangle. 
\end{equation}
Here $\langle \cdots \rangle$ stands for the statistical average. 
We also introduce the orientational order parameter for the stripe order: 
\begin{equation}
O_{hv} = \left| \frac{n_h - n_v}{n_h + n_v} \right|, 
\label{Ohv}
\end{equation}
where
\begin{eqnarray}
n_h & = & \sum _{i} \left( 1 -  \vector{S} _i. \cdot \vector{S} _{i+\vector{e}_x} \right)  \;\;\; \label{nh} \\
{\rm and} \;\;\;  n_v & = &\sum _{i} \left( 1 -  \vector{S} _i. \cdot \vector{S} _{i+\vector{e}_y} \right) \label{nv}
\end{eqnarray}
as defined in Ref.~\cite{Whitehead}. 

We identify second-order phase transitions by evaluating the fourth-order Binder cumulants~\cite{Binder} for $M_{xy}$ and $O_{hv}$ defined by  
\begin{equation}
{U}_4 = 1-\frac{\left< M_{xy} ^4 \right> }{3 \left< M_{xy} ^2 \right> ^2} , \label{MBinder}
\end{equation}
and
\begin{equation}
\tilde{U}_4 = 1-\frac{\left< O_{hv} ^4 \right> }{3 \left< O_{hv} ^2 \right> ^2} , \label{OBinder} 
\end{equation}
respectively, and first-order phase transitions by investigating hysteresis loops of the order parameters for strong hysteresis or energy histograms for weak hysteresis.

As mentioned in the introduction, naive MC methods require $O(N^2)$ simulation time, and thus we adopt the $O(N)$ SCO algorithm~\cite{Sasaki} for the MC simulation in the present study. 
The SCO algorithm is based on the stochastic potential switching (SPS) algorithm with $O(N)$ switching time for long-range interactions~\cite{Mak1,Mak2}, and was first introduced to the dipolar Heisenberg model without uniaxial anisotropy. 
However, when the uniaxial anisotropy is large such as the dipolar Ising model, the SPS procedure does not work efficiently for short-range contribution. Then, all the interactions up to a certain range should be taken into account in the conventional algorithm, and the SPS procedure should only be applied outside of that range. Here we tune the range in the same way as in our previous study~\cite{Komatsu}. 
This modification provides enough efficiency in the MC sampling. We use $0.5\times10^5$--$1.0 \times 10^5$ MC steps for equilibration and $1.5\times10^5$--$4\times10^5$ MC steps for measurement at each temperature (or $\eta$) during gradual change of temperature (or $\eta$).

\begin{figure}[t!]
\begin{center}
\includegraphics[width = 4.25cm]{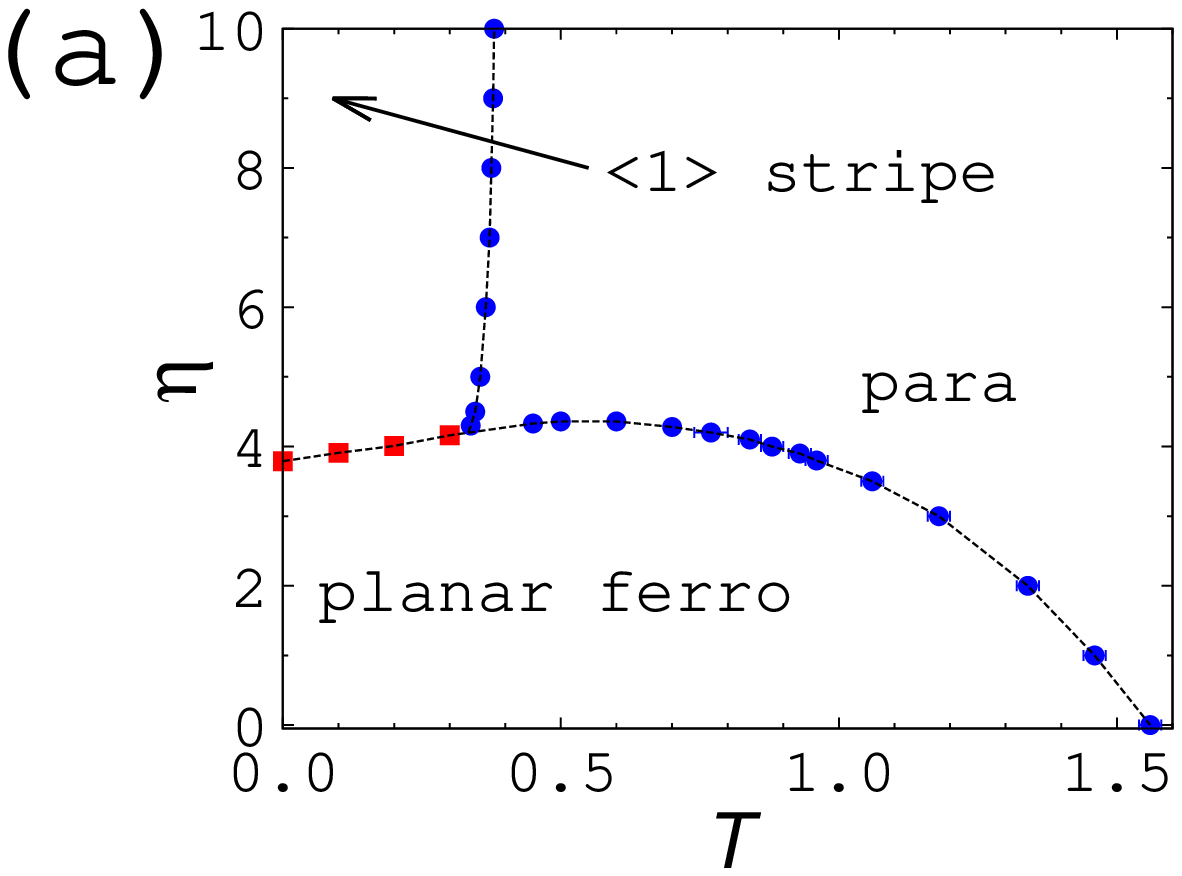}
\includegraphics[width = 4.25cm]{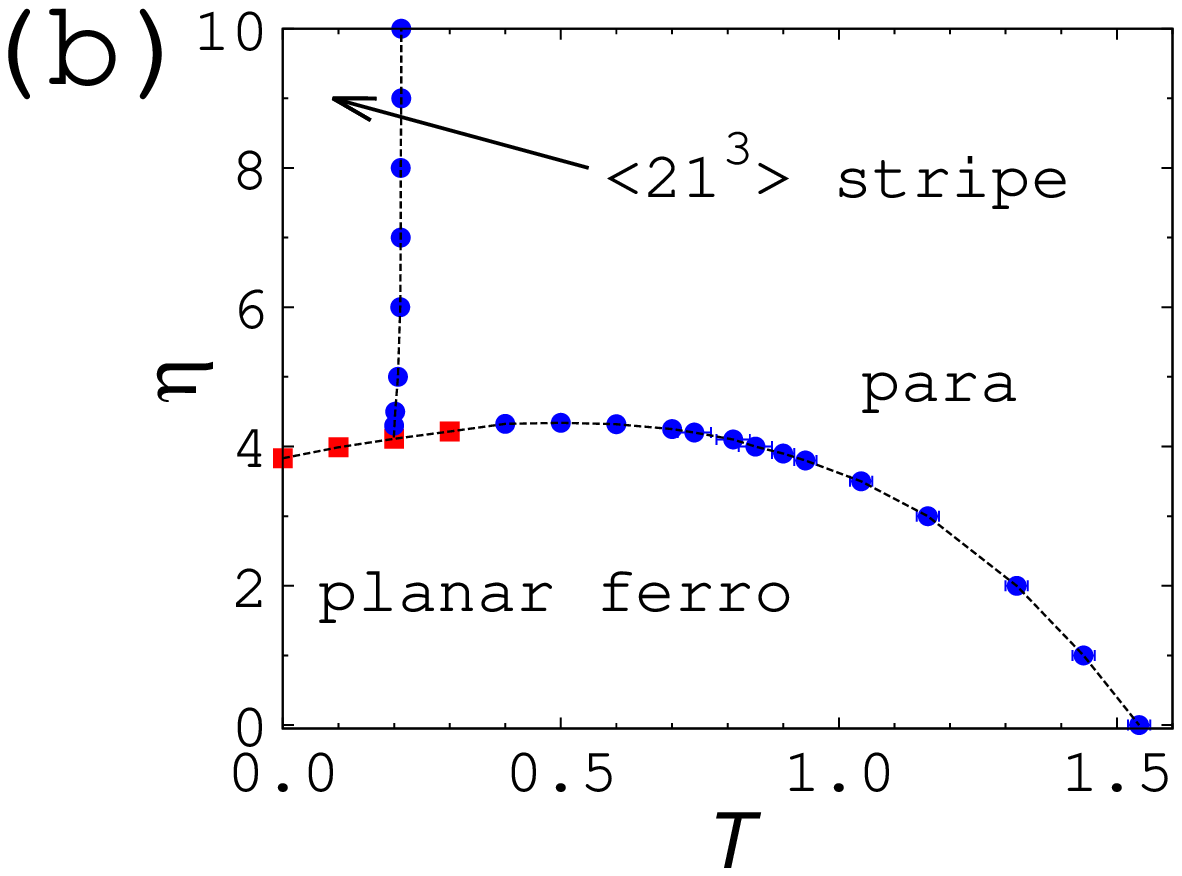}

\includegraphics[width = 4.25cm]{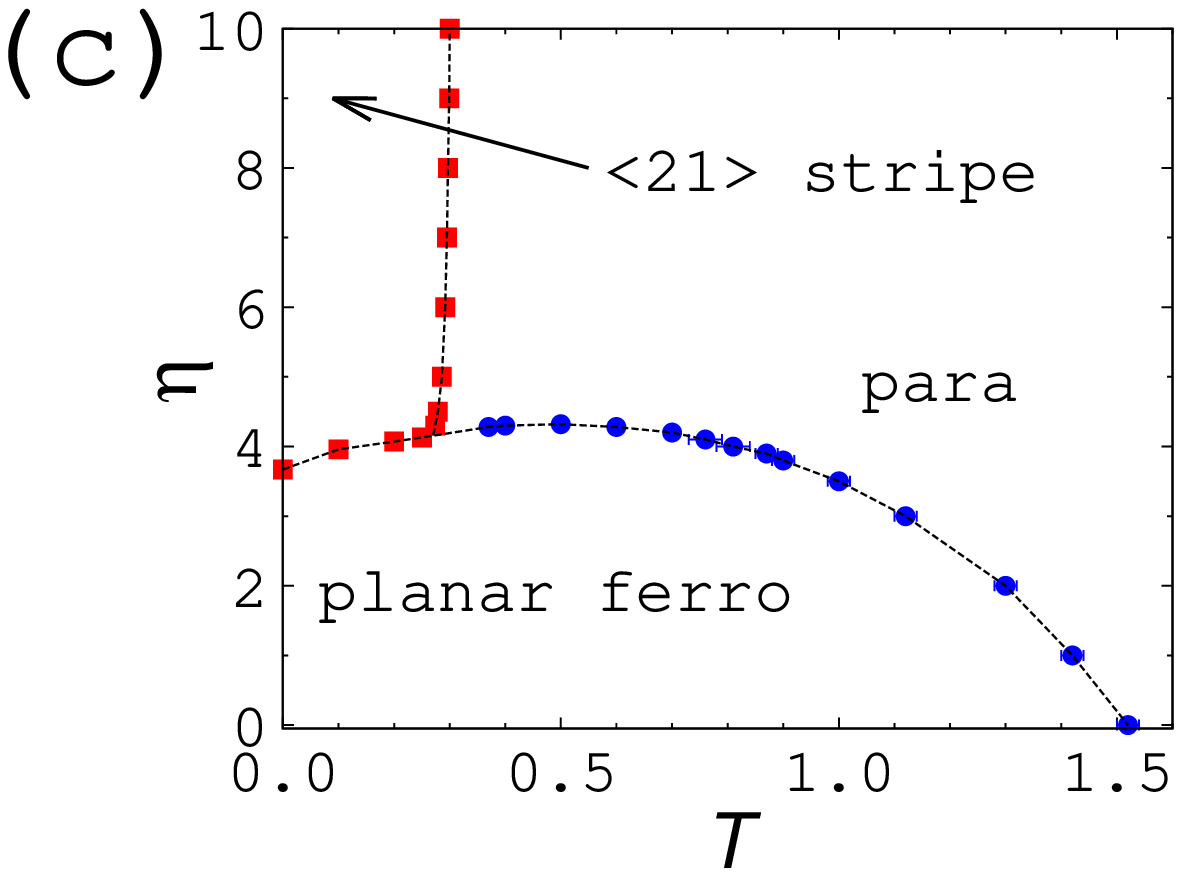}
\includegraphics[width = 4.25cm]{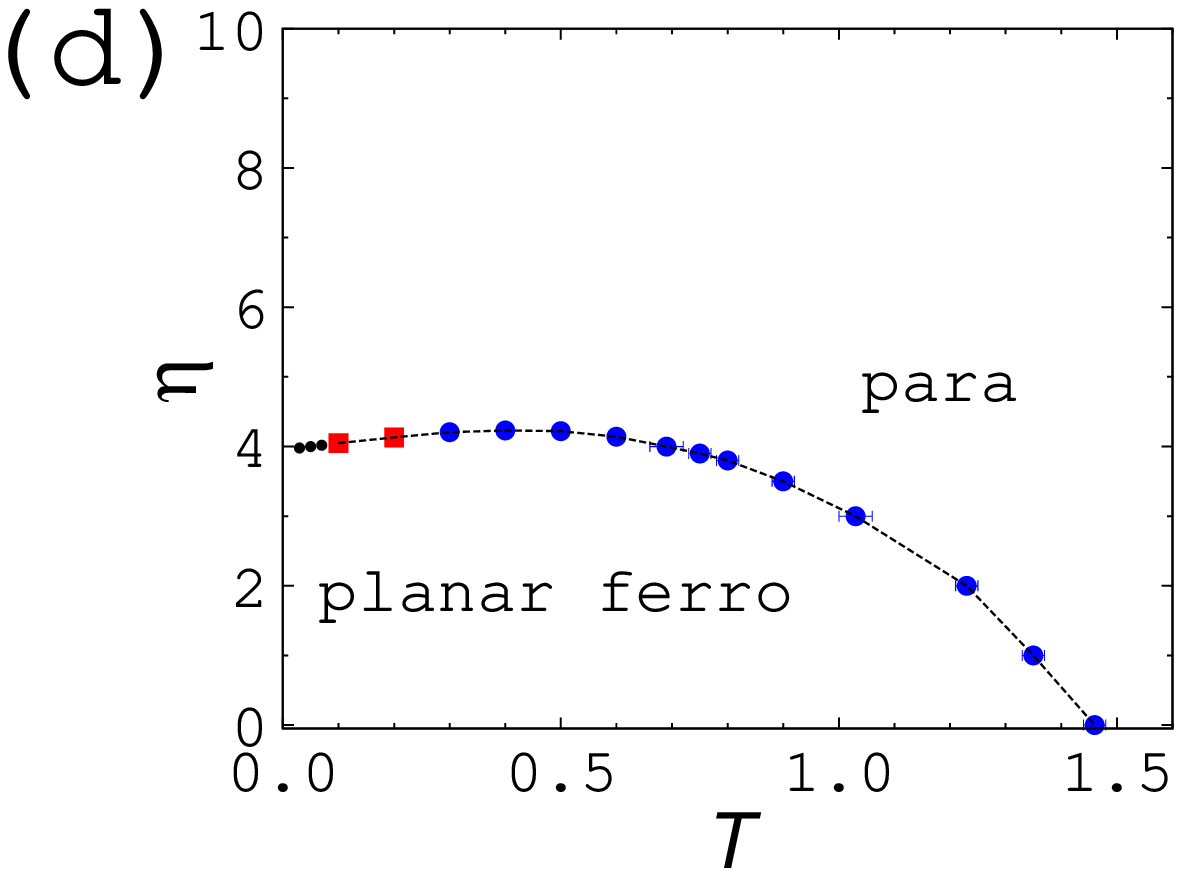}
\end{center}
\caption{$\eta$--$T$ phase diagrams at (a) $H =0$, (b) $H=1.3$, (c) $H = 2$, and (d) $H=3.2$. The red squares and blue circles are first- and second-order transition points, respectively.}
\label{PD}
\end{figure}

\begin{figure}[t!]
\begin{center}
\includegraphics[width = 4.25cm]{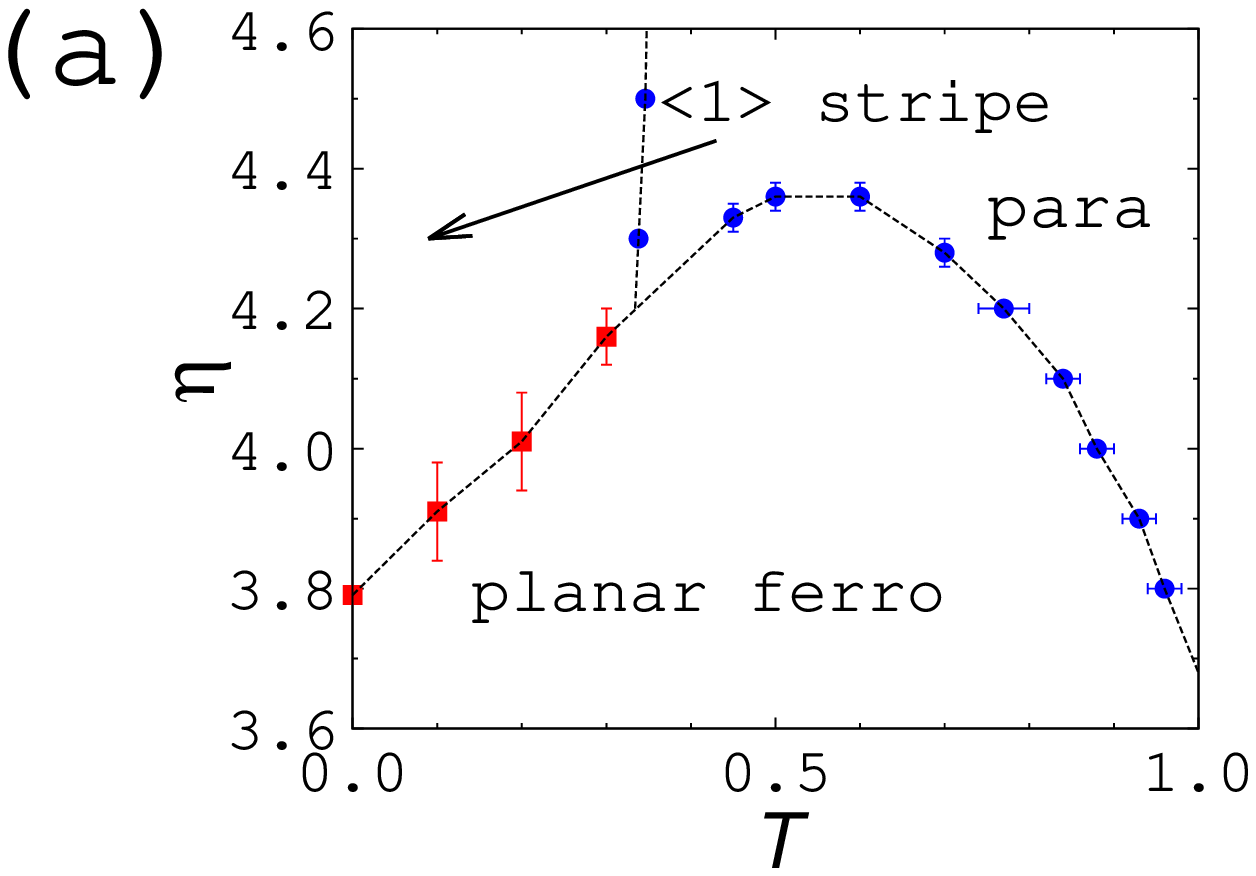}
\includegraphics[width = 4.25cm]{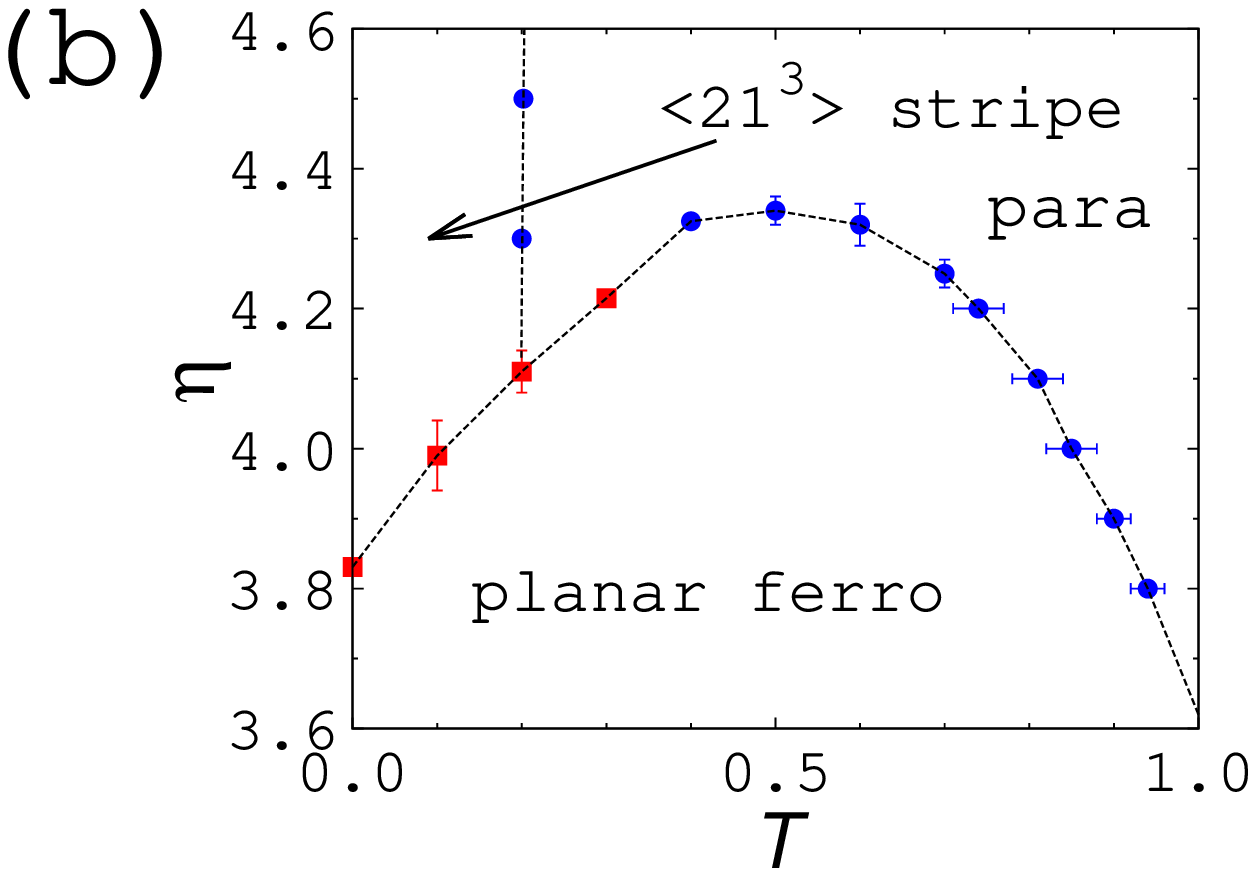}

\includegraphics[width = 4.25cm]{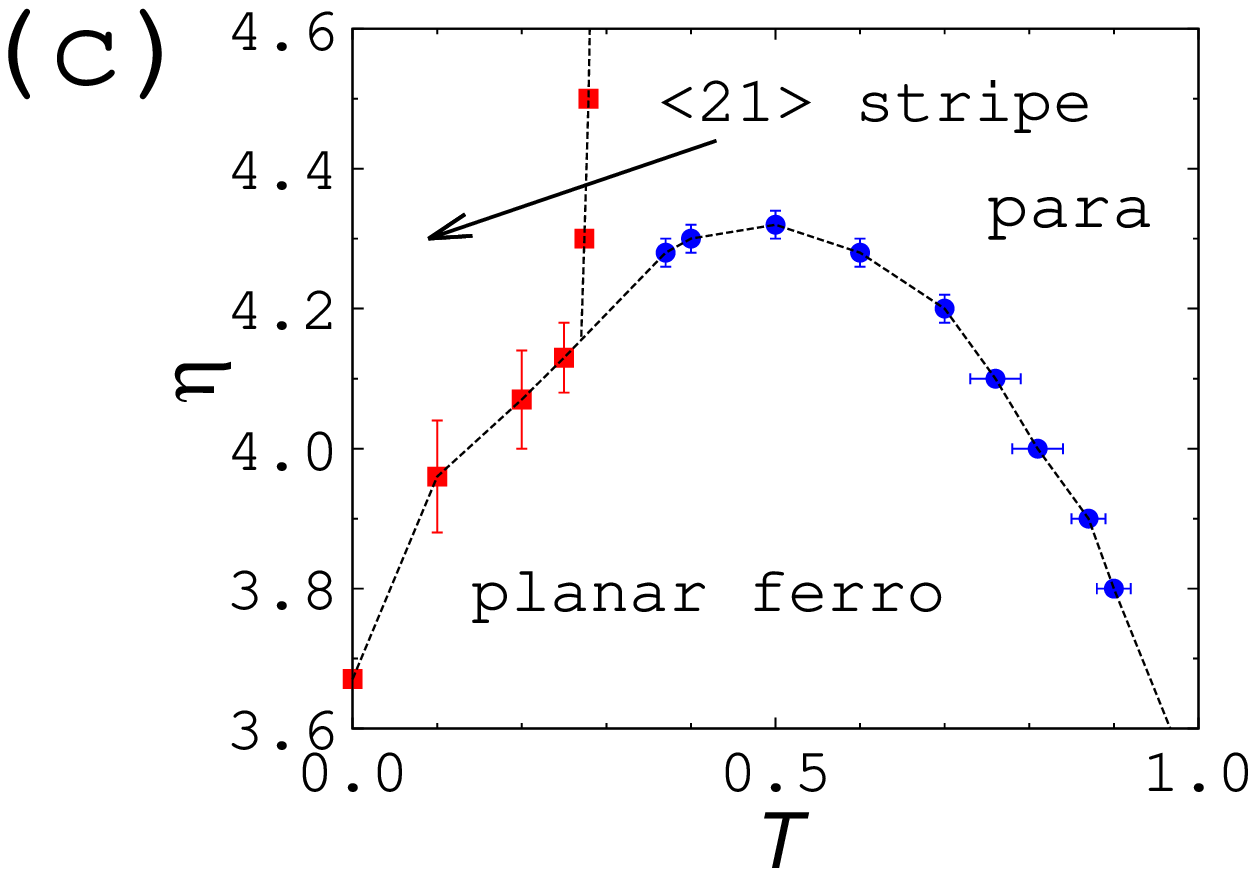}
\includegraphics[width = 4.25cm]{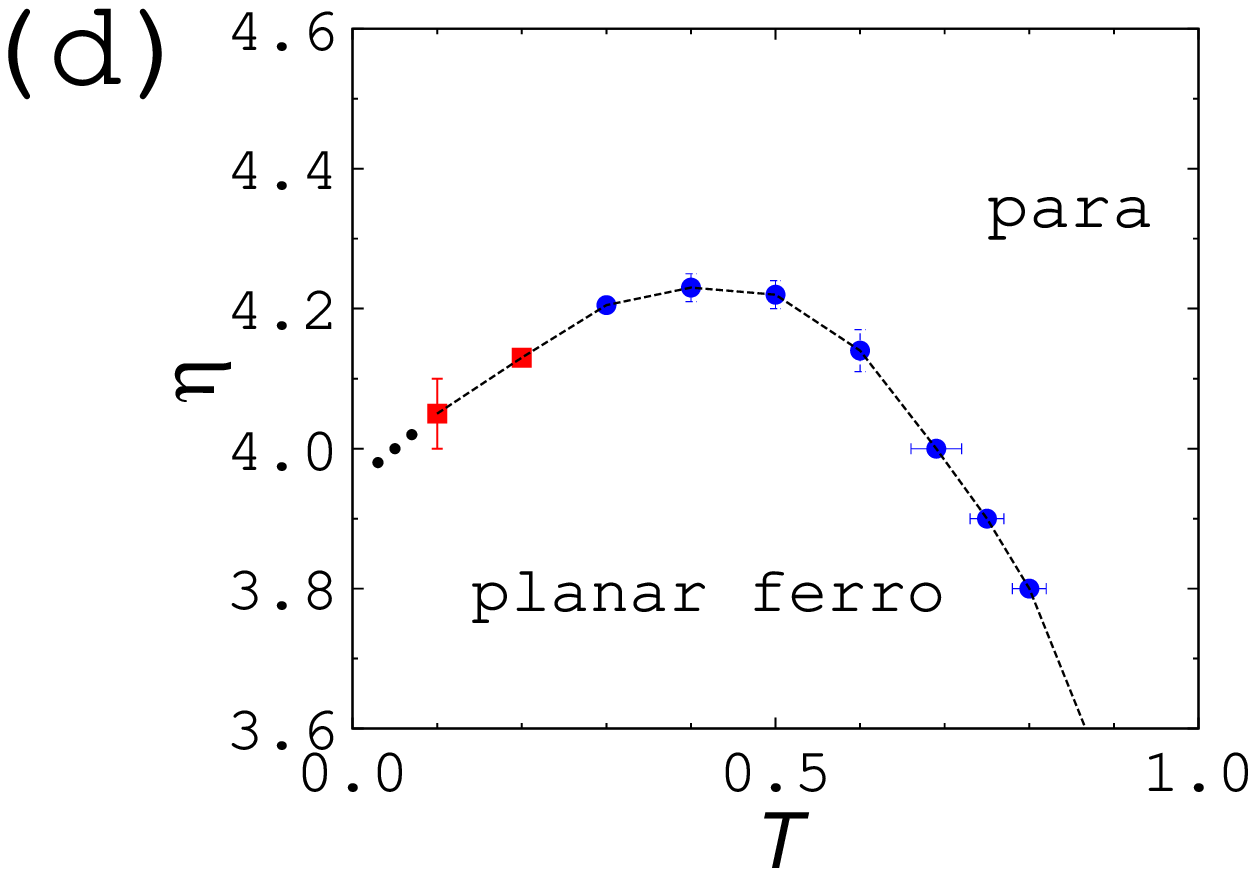}
\end{center}
\caption{Magnified phase diagrams at (a) $H=0$, (b) $H=1.3$, (c) $H=2$, and (d) $H=3.2$. The red squares and blue circles denote first- and second-order transition points, respectively. The error bars of the first-order transition points for the SR transition line are estimated from the widths of the hysteresis loops of $O_{hv}$ and $M_{xy}$.
}
\label{PD_magnified}
\end{figure}

\begin{figure}[btp!]
\begin{center}
\includegraphics[width = 8.5cm]{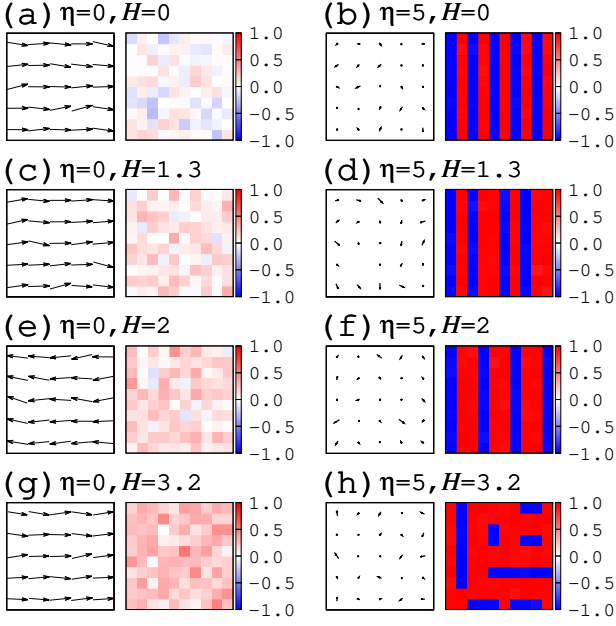}
\end{center}
\caption{Snapshots of typical spin configurations at a low temperature ($T=0.1$) for different $\eta$ and $H$. 
In each figure ((a)--(h)), arrows (left panel) denote vectors $(S_i ^x, S_i ^y)$ in a block of 5 $\times$ 5 spins. 
The graduation of the color (right panel) indicates the magnitude of $S_i ^z$ (${\rm red}=1$, ${\rm white}=0$, ${\rm blue}=-1$ in a block of 10 $\times$ 10 spins. In the left panels of (a), (c), (e), and (g), spins are mainly oriented along the $x$ direction ($x$ and $y$ directions are equivalent), which suggests stability of the planer F phase for any fields at $\eta=0$. The right panels of (b), (d), and (f) visualize the stripe patterns: $\langle 1 \rangle$, $\langle 21^3 \rangle$ and $\langle 21  \rangle$ at $H=0$, 1.3, and 2, respectively, and that of (h) shows no stripe order at $H=3.2$.}
\label{Snapshots}
\end{figure}

\begin{figure}[thbp!]
\begin{center}
\includegraphics[width = 4.25cm]{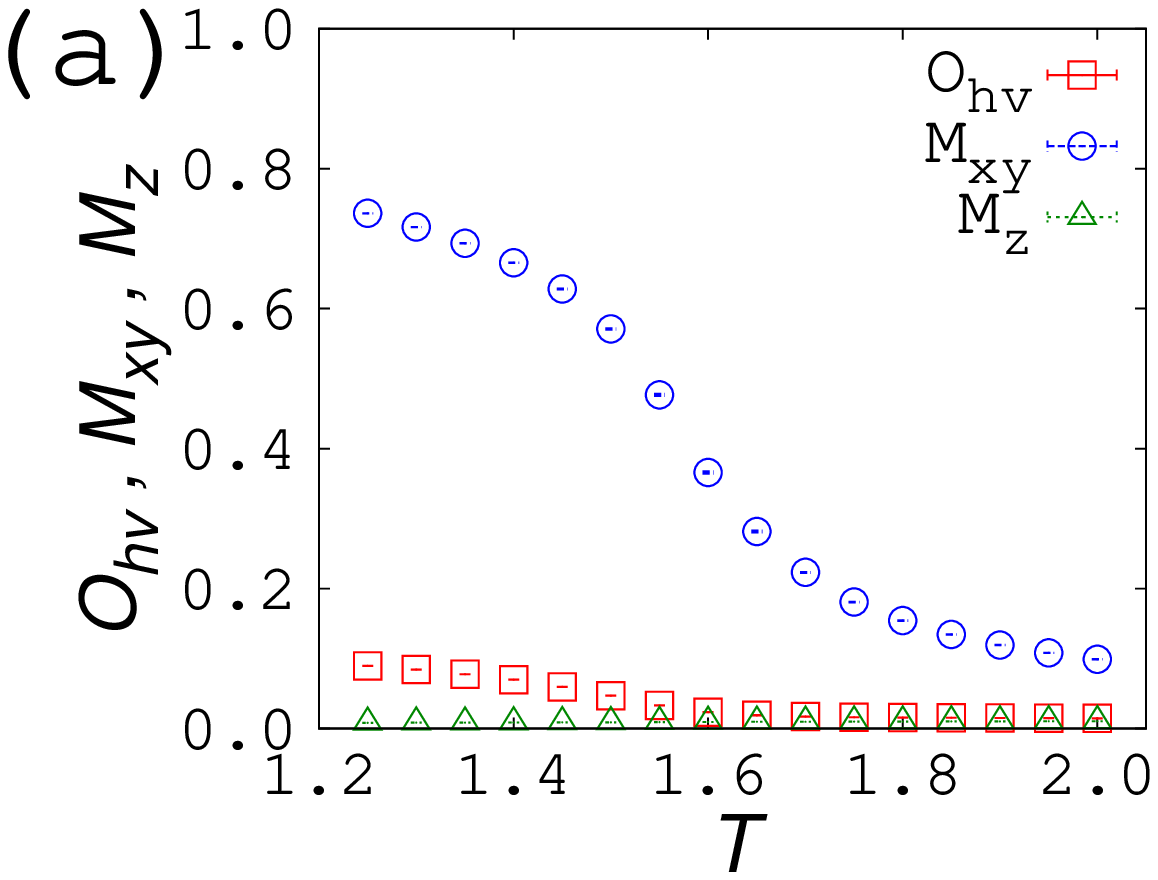}
\includegraphics[width = 4.25cm]{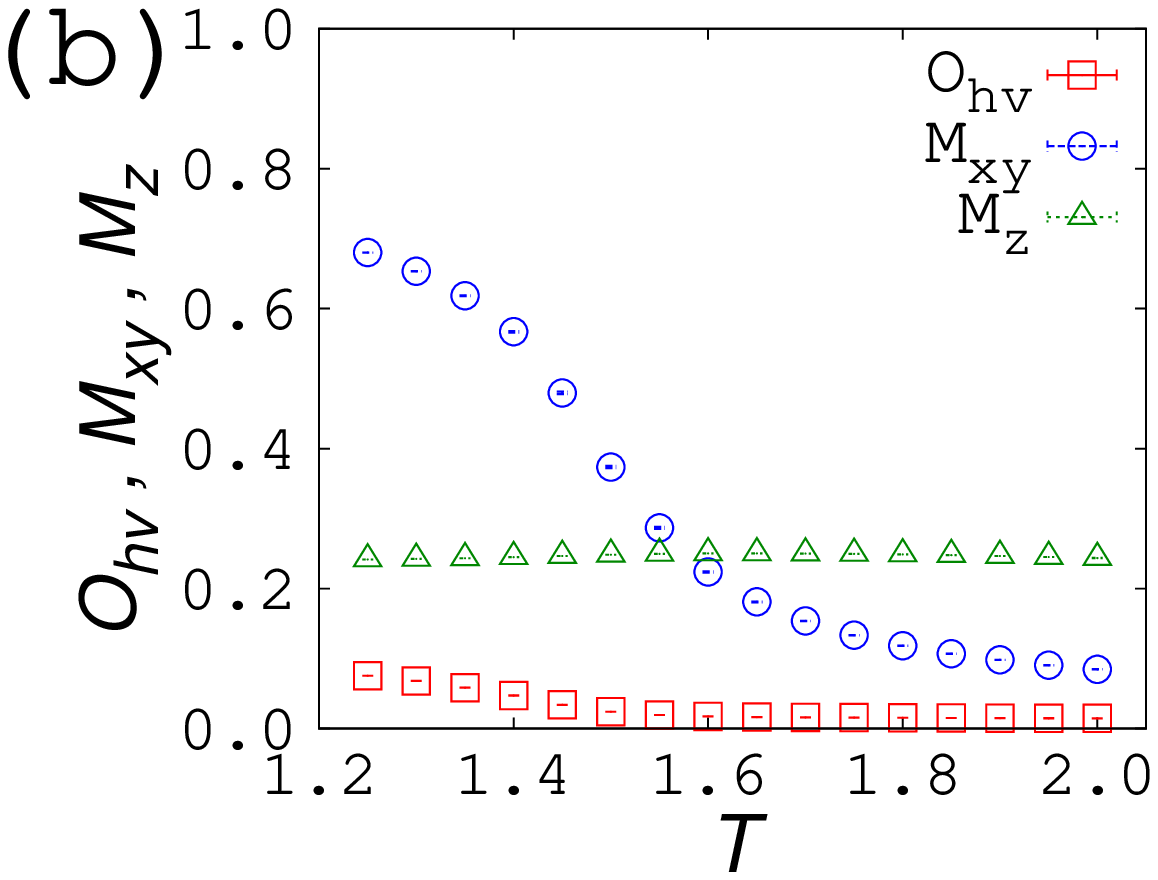}

\includegraphics[width = 4.25cm]{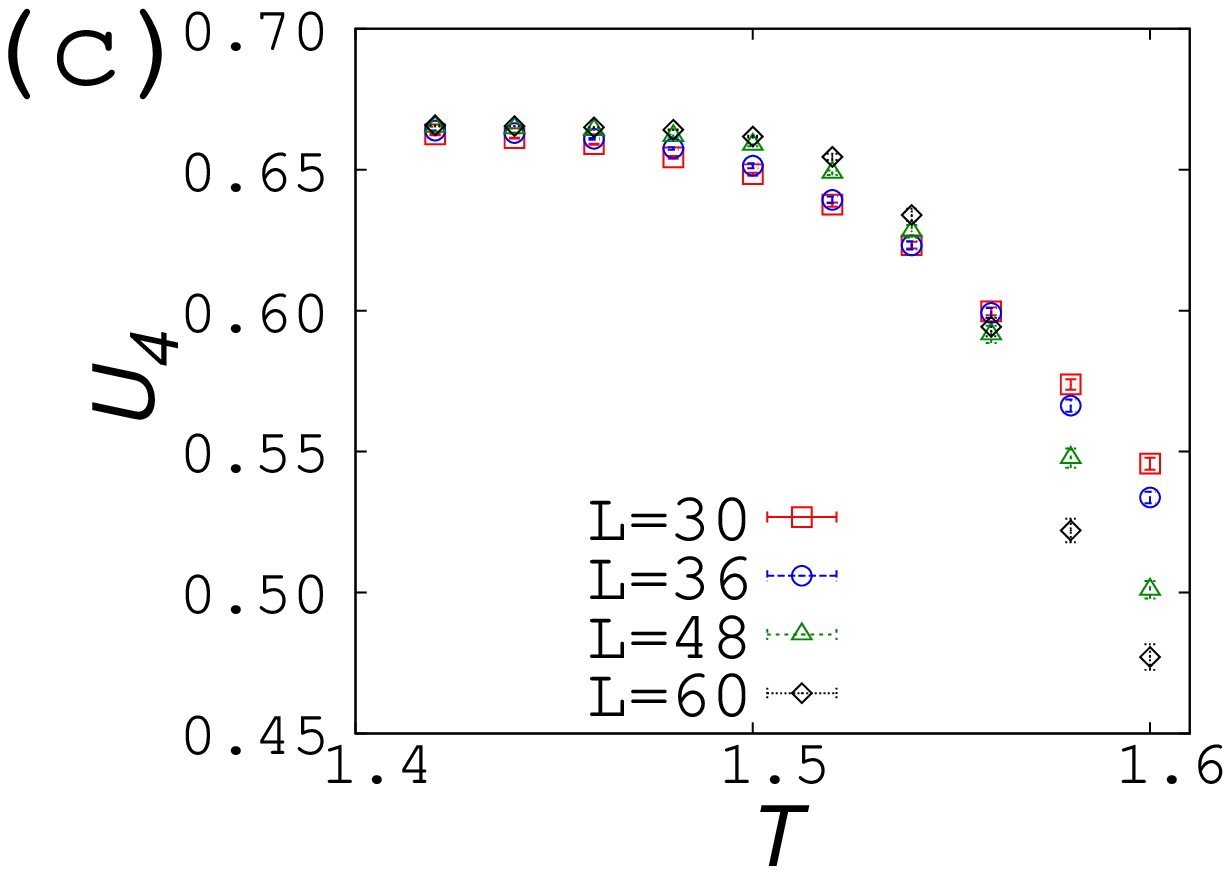}
\includegraphics[width = 4.25cm]{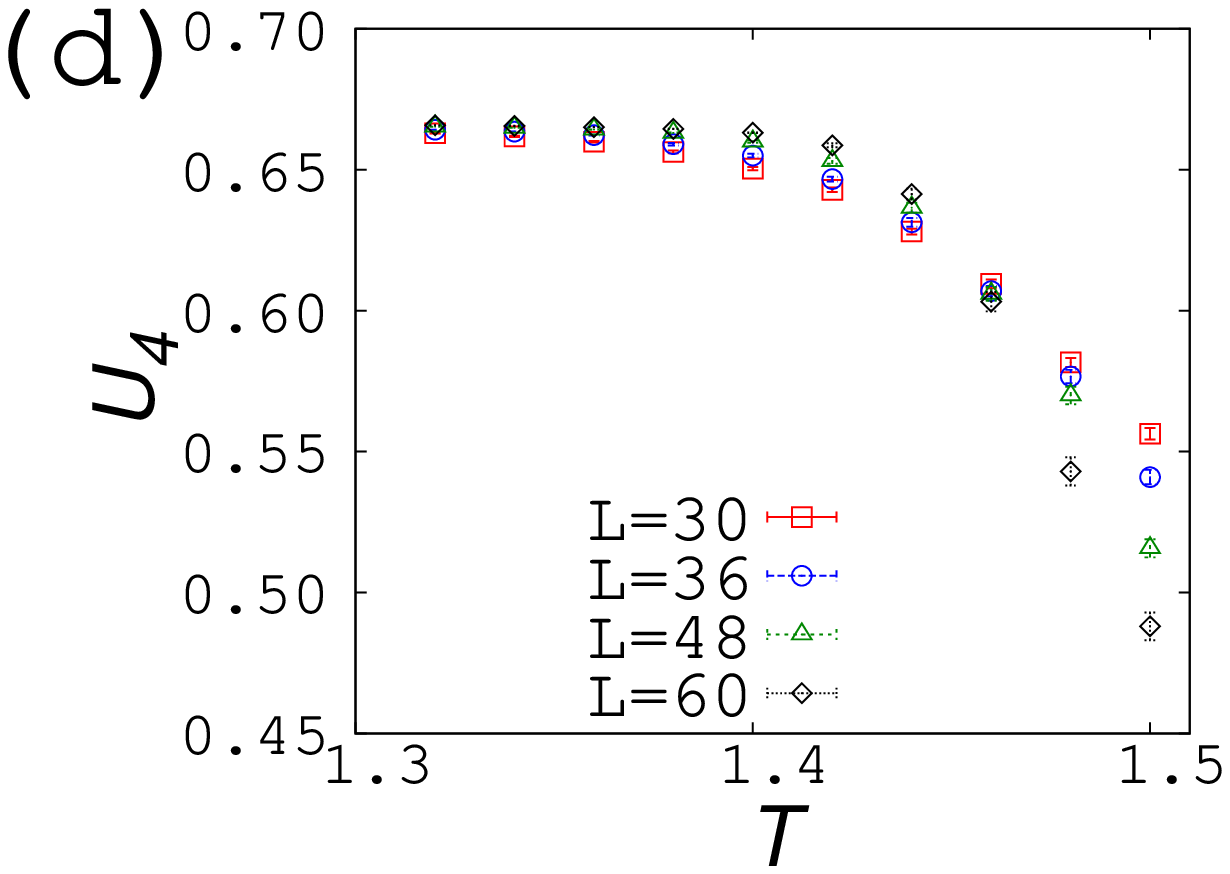}
\end{center}
\caption{Temperature dependence of $O_{hv}$ (red squares), $M_{xy}$ (blue circles), and $M_z$ (green triangles) for $L=30$ at (a) $H=0$ and $\eta = 0$, and (b) $H=3.2$ and $\eta = 0$, and that of $U_4$ at (c) $H=0$ and $\eta = 0$ and at (d) $H=3.2$ and $\eta = 0$ for $L=30$, $36$, $48$ and $60$.}
\label{T_vs_OP_eta0}
\end{figure}

\section{Monte Carlo study}
\label{MC_study}

\subsection{Overview of the phase diagrams}
\label{overview}

First we overview the phase diagrams obtained by the MC method in the present study.         
The $\eta$--$T$ phase diagrams at $H=0$, $1.3$, $2$ and $3.2$ are given in Figs.~\ref{PD}(a)--\ref{PD}(d), and the magnified ones around the triple and/or tricritical points are shown in Figs.~\ref{PD_magnified}(a)--\ref{PD_magnified}(d). 
Snapshots of typical spin configurations at a low temperature ($T=0.1$) are illustrated in Figs.~\ref{Snapshots}(a)--\ref{Snapshots}(h). 
At $H=0$ (Fig.~\ref{PD}(a),  Fig.~\ref{PD_magnified}(a)) planar F, SO  $\langle 1 \rangle$, and P phases appear.  
At relatively low fields, similarly three phases exist but the SO phase varies the stripe pattern as $\langle 1 \rangle \rightarrow \langle 21^3 \rangle \rightarrow  \langle 21  \rangle$ with rising field (Figs.~\ref{PD}(b) and \ref{PD}(c),  Figs.~\ref{PD_magnified}(b) and ~\ref{PD_magnified}(c)). 
At relatively high fields, e.g., $H=3.2$, the SO phase disappears but the planar F phase still remains (Fig.~\ref{PD}(d),  Fig.~\ref{PD_magnified}(d)).

\subsection{Phase transition between the paramagnetic and planar ferromagnetic phases}
\label{PF_and _D}

Hereafter we discuss details of the phase transitions. 
First we focus on the transition between the P and planar F phases at relatively small $\eta$. 
The temperature dependence of $O_{hv}$, $M_{xy}$, and $M_z$ at $\eta=0$ for $H=0$ and $H=3.2$ is shown in Figs.~\ref{T_vs_OP_eta0}(a) and \ref{T_vs_OP_eta0}(b), respectively. 

The transverse magnetization $M_{xy}$ grows below $T \approx 1.6$ at $H=0$ and $T \approx 1.5$ at $H=3.2$, and the stripe order $O_{hv}$ is almost zero, which indicates the phase transition between the P and planar F phases. It is noting that $M_z \ne 0$ at $H=3.2$ (actually also at $H=1.3$ and $2$) and the spins in the planar F phase are tilted in the $z$ direction. 
The temperature dependence of $U_4$ for $L=30$, 36, 48, and 60 at $\eta=0$ is given for $H=0$ and $H=3.2$ in Figs.~\ref{T_vs_OP_eta0}(c) and \ref{T_vs_OP_eta0}(d), respectively. The $U_4$ curves cross at $T = 1.55$ and $T=1.46$ in Figs.~\ref{T_vs_OP_eta0}(c) and \ref{T_vs_OP_eta0}(d), respectively, which suggests second-order phase transitions. In the same way, we identify the second-order phase transition points between the P and planar F phases in the phase diagrams (Figs.~\ref{PD} and \ref{PD_magnified}), denoted by blue circles.

\begin{figure}[thbp!]
\begin{center}
\includegraphics[width = 4.25cm]{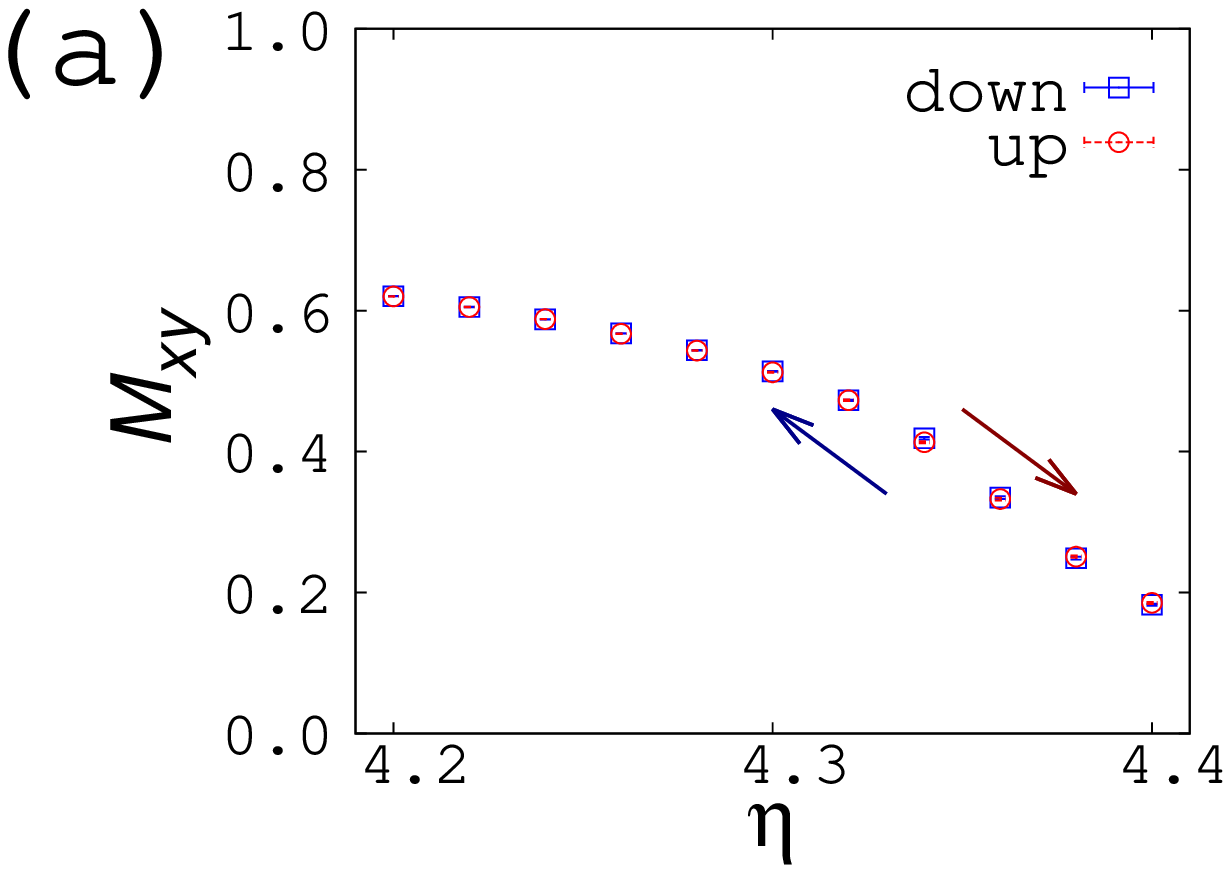}
\includegraphics[width = 4.25cm]{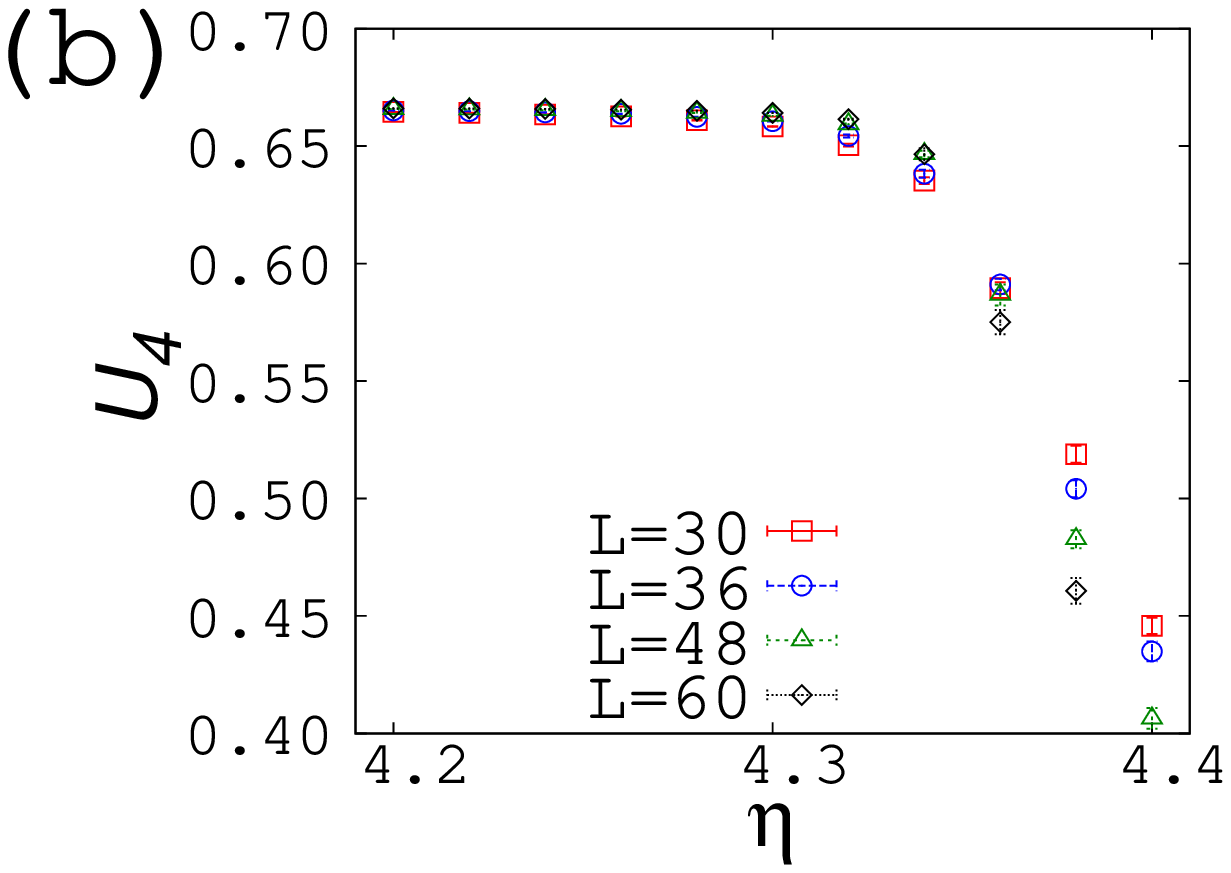}
\end{center}
\caption{$\eta$ dependence at $H=0$ and $T=0.5$ of (a) $M_{xy}$ for $L=30$ and (b) $U_4$ for $L=30$, $36$, $48$ and $60$.}
\label{eta_vs_Mxy_U4_H0_T0.5}
\end{figure}

\begin{figure}[thbp!]
\begin{center}
\includegraphics[width = 6.0cm]{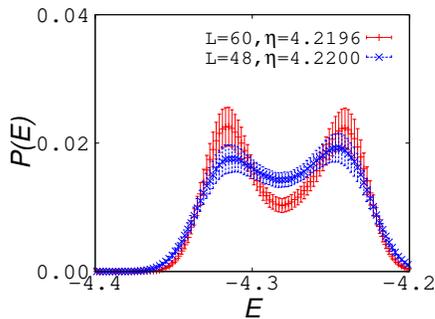}
\end{center}
\caption{Energy histogram $P(E)$ at $H=1.3$ and $T=0.3$ near the transition point between the paramagnetic and planar ferromagnetic phases (see also Figs.~\ref{PD}(b) and \ref{PD_magnified}(b)). $P(E)$'s for $L=60$ at $\eta = 4.2196$ and for $L=48$ at $\eta = 4.2200$ show double peaks and the peaks become sharper for larger $L$.}
\label{Hist_H1.3_T0.3}
\end{figure}

\begin{figure}[thbp!]
\begin{center}
\includegraphics[width = 4.25cm]{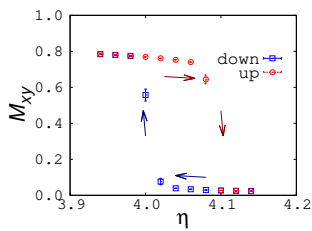}
\end{center}
\caption{$\eta$ dependence of $M_{xy}$ at $H=3.2$ and $T=0.1$ for $L=30$.}
\label{eta_vs_Mxy_H3.2_T0.1}
\end{figure}

Next we investigate the phase boundaries around the triple and/or tricritical points. 
The $\eta$ dependence of $M_{xy}$ and $U_4$ at $H=0$ and $T=0.5$ (Fig.~\ref{PD}(a) and  Fig.~\ref{PD_magnified}(a)) are depicted in Figs.~\ref{eta_vs_Mxy_U4_H0_T0.5}(a) and \ref{eta_vs_Mxy_U4_H0_T0.5}(b), respectively. 
The transverse magnetizations $M_{xy}$ with increasing and decreasing $\eta$ overlap well, and the corresponding $U_4$ curves for different system sizes cross at $\eta=4.35$, and thus the transition is of second order.

For $H=1.3$, we find first-order transition points in the vicinity of the triple point on the P--planer F transition line (Fig.~\ref{PD} (b) and Fig.~\ref{PD_magnified} (b)), and there exists a tricritical point on the line. 
Figure~\ref{Hist_H1.3_T0.3} illustrates energy histograms at $H=1.3$ and $T=0.3$. Around the transition point $\eta \approx 4.22$, the histograms show double peaks with a little size dependence and the peaks become sharper for larger $L$, which is an evidence for the first-order transition. For $H=2$, we obtain the second-order P--planar F transition line,  similar to that for $H=0$.

It should be noted that any SO phase does not exist at $H=3.2$, and that at low temperatures, a first-order-transition line between the P and planar F phases appears (Fig.~\ref{PD} (d) and Fig. \ref{PD_magnified} (d)), which is confirmed by strong hysteresis of $M_{xy}$ around the line (see Fig.~\ref{eta_vs_Mxy_H3.2_T0.1} for $T=0.1$). It terminates at a tricritical point. 
We find that the planar F state is robust against the external field, which 
is probably because the $z$ component of the spins in the planar F phase can change continuously as the field varies. 
Here we find that in all cases the maximum point of $\eta$ of the transition line between the planar F and P phases exists at an intermediate temperature. 
A similar feature was found in the $\eta$--$T$ diagram for $\delta=3$ and $H=0$ by Carubelli {\it et al}.~\cite{Carubelli}.

The existence of the tricritical point on the P--planar F transition line is nontrivial. We observe the tricritical point on the transition line for $H=1.3$ and $H=3.2$ but cannot for $H=0$ and $H=2$ within the accuracy of the present simulation. 
On the other hand, in the case of $\delta=3$ ~\cite{MacIsaac2}, the tricritical point was reported for $H=0$, and we do not rule out the possibility of the tricritical point close to the triple point for $H=0$ and $H=2$ for $\delta=1$.

The existence of the tetragonal liquid (TL) phase between the P and SO phases has been pointed out theoretically~\cite{Abanov,Booth} and experimentally~\cite{Vaterlaus}. 
Booth {\it et al.}~\cite{Booth} reported that in the temperature dependence of the specific heat, a broad shoulder with a peak follows the sharp peak as the temperature is raised for $\delta=4.45$ in the Ising limit. They suggested that the former originates from the transition between the TL and P phases and the latter from that between the SO and TL phases. However, the TL phase has no long-range order and the broad shoulder might indicate a crossover rather than the phase transition. 
Furthermore they also showed the broad shoulder becomes much smaller for $\delta=3$. Here we could not detect any signs of the TL phase for $\delta=1$. 

\begin{figure}[th]
\begin{center}
\includegraphics[width = 4.25cm]{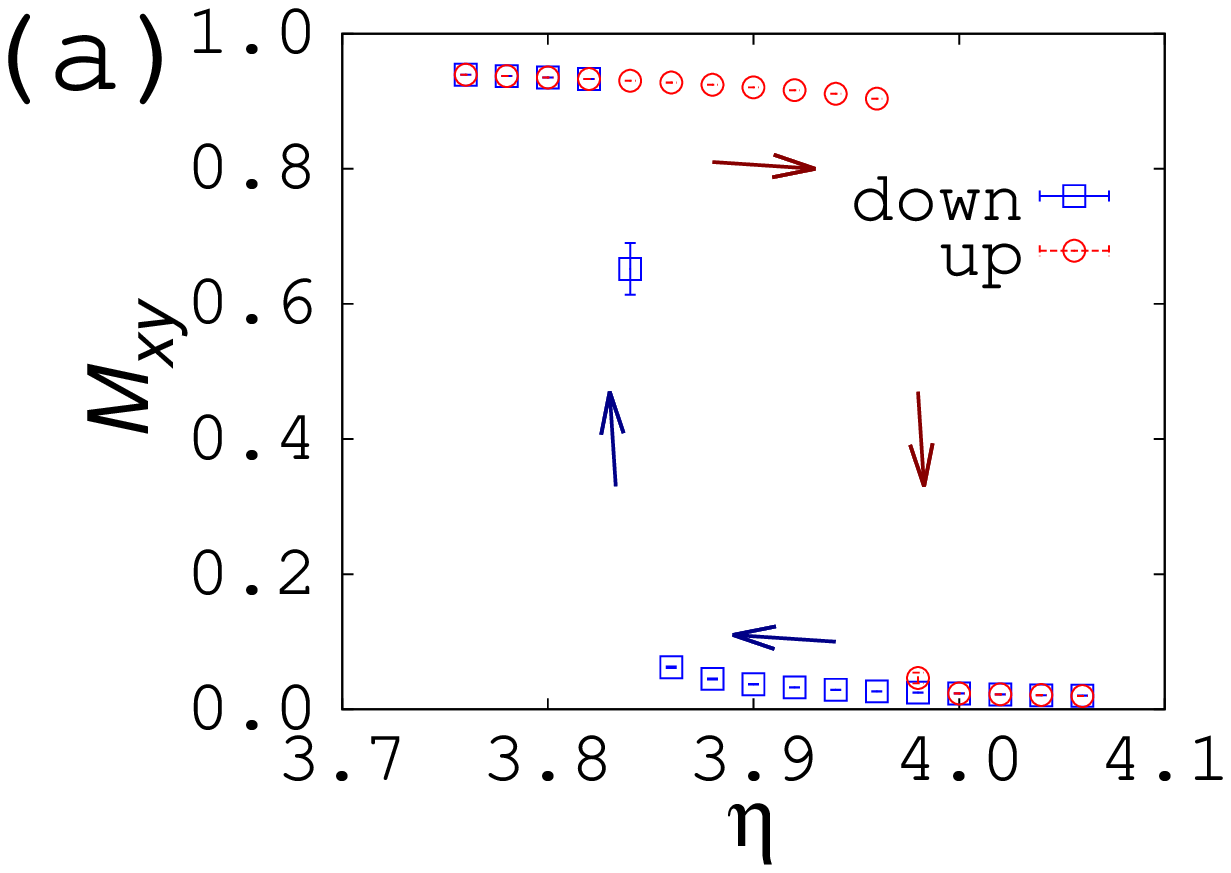}
\includegraphics[width = 4.25cm]{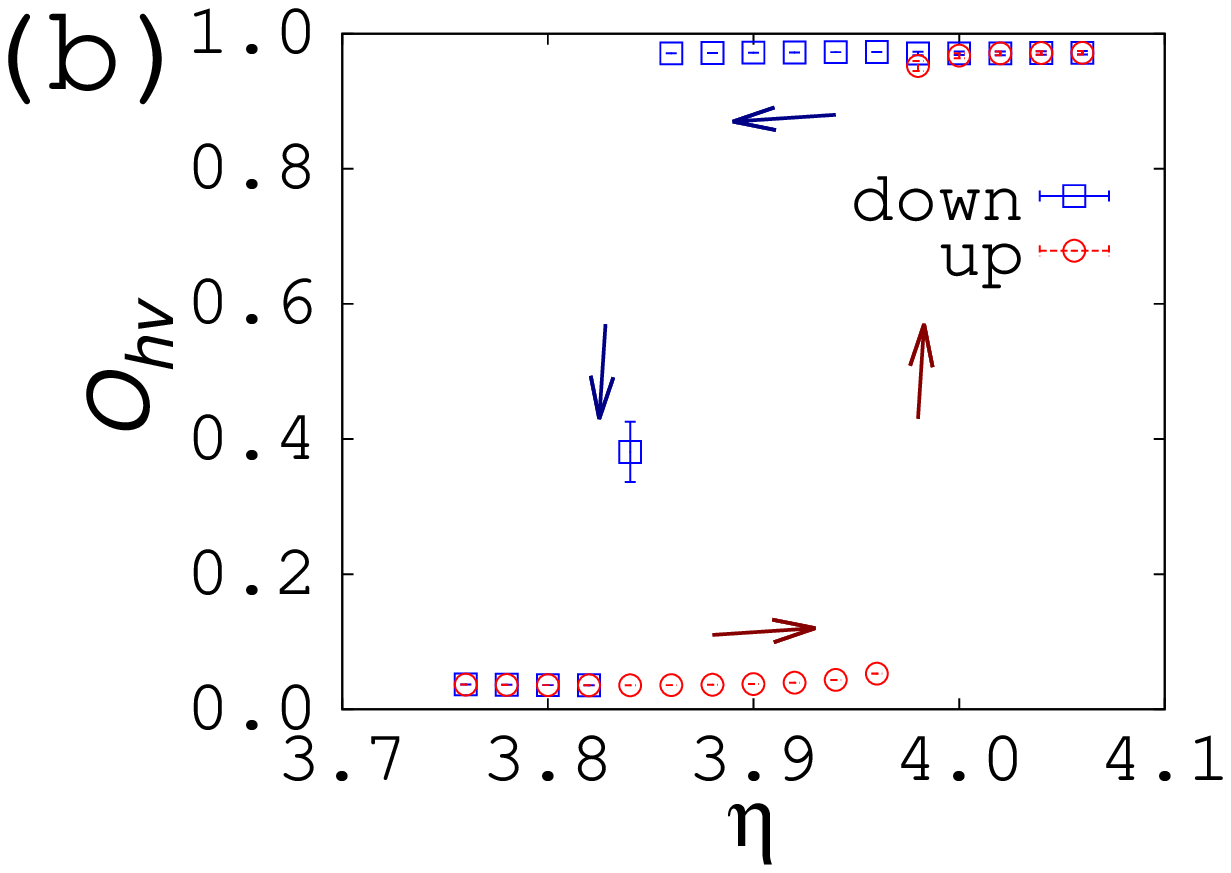}
\end{center}
\caption{$\eta$ dependence of (a) $M_{xy}$ and (b) $O_{hv}$ at $H=0$ and $T=0.1$ for $L=30$. Large hysteresis loops are observed around the SR transition point.}
\label{eta_vs_Mxy_H0_T0.1}
\end{figure}

\begin{figure}[th]
\begin{center}
\includegraphics[width = 4.25cm]{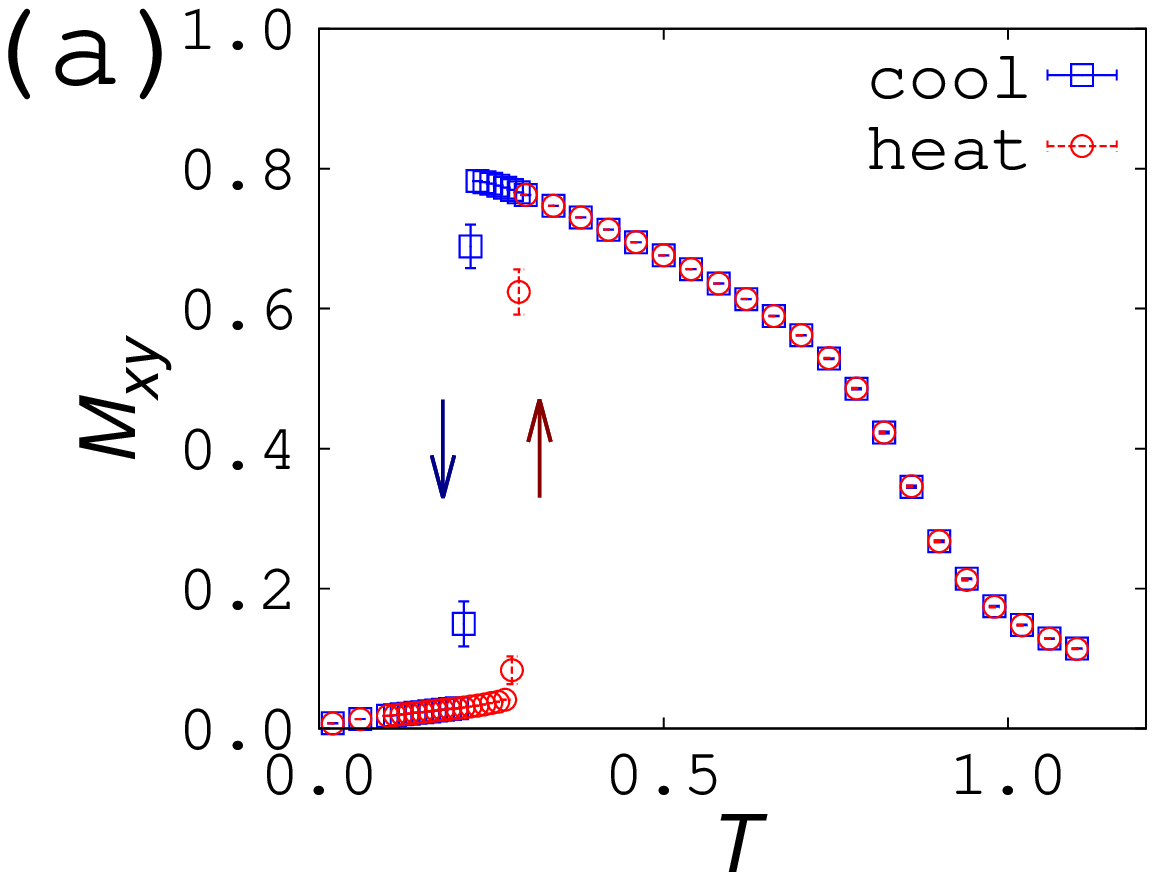}
\includegraphics[width = 4.25cm]{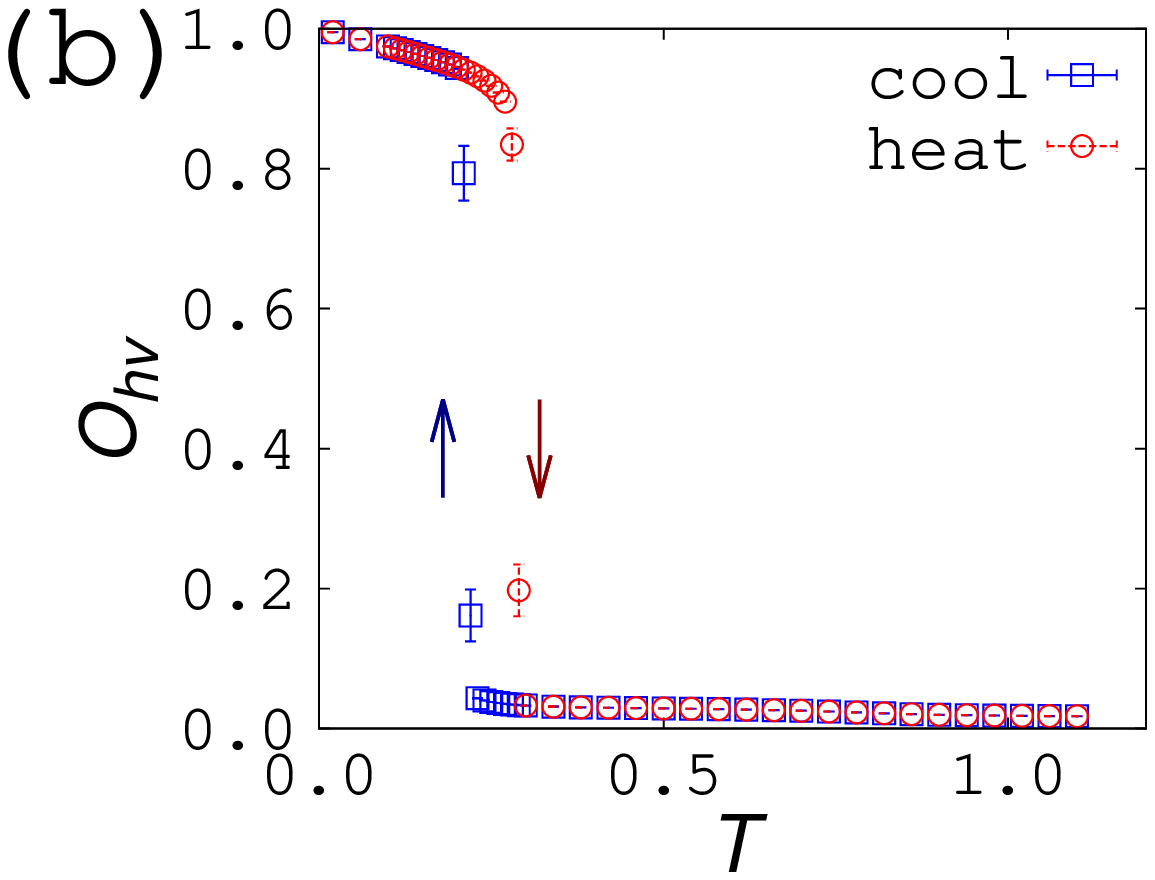}
\end{center}
\caption{Temperature dependence of (a) $M_{xy}$ and (b) $O_{hv}$ at $H=0$ and $\eta=4.1$ for $L=30$. 
Hysteresis loops are observed around the SR transition point.}
\label{reentrant1}
\end{figure}

\subsection{Spin reorientation transition between the planar ferromagnetic and stripe-ordered phases}
\label{SR-transition}

Strong hysteresis of the order parameters is observed around the boundary between the planar F and SO phases with and without field. 
For example, at $H=0$ and $T=0.1$, we see large hysteresis loops of $M_{xy}$ and $O_{hv}$ along the $\eta$ direction in Figs.~\ref{eta_vs_Mxy_H0_T0.1}(a) and \ref{eta_vs_Mxy_H0_T0.1}(b), respectively, and also at $H=0$ and $\eta=4.1$, along the $T$ direction in Figs.~\ref{reentrant1}(a) and \ref{reentrant1}(b), respectively. 
The first-order transition for SR is consistent with the previous studies for different $\delta$~\cite{Pescia,Hucht,MacIsaac1,MacIsaac2,Carubelli,Santamaria}. 
In the same way, the first-order transition points for SR are identified in the phase diagrams with and without field. 
Here the transition points at $T=0$ in Figs.~\ref{PD_magnified}(a), \ref{PD_magnified}(b), and \ref{PD_magnified}(c) are evaluated by comparing the energies for the perfectly-ordered planar F and SO phases.

The existence of a canted SO phase between the planar F and SO phases was indicated for $h_i \ge 3$ ($\delta>2$) at $H=0$ in the ground state ($T=0$)~\cite{Pighin2}. Indeed, for $\delta=4.45$~\cite{Whitehead} and $\delta=6$~\cite{Pighin} at $H=0$, canted SO phases between the planar F and SO phases were reported at low temperatures. In the present study the maximum stripe width in the SO phases is $h_i=2$, and any canted SO phase does not exist for $\delta=1$ with and without field. 

We find that for $\delta=1$, $\frac{d\eta}{dT}>0$ is realized in the SR transition at $H=0$ in  Fig.~\ref{PD_magnified}(a). 
This result for $\delta=1$ has the same tendency as that for $\delta=3$ by Carubelli {\it et al}.~\cite{Carubelli} and does not as that by MacIsaac {\it et al}.~\cite{MacIsaac2} 
We also find that $\frac{d\eta}{dT}>0$ in the SR transition holds for finite fields. 
It is worth noting that instead of the SR transition line, $\frac{d\eta}{dT}>0$ is realized for the P--planar F line at $H=3.2$.

\begin{figure}[tb]
\begin{center}
\includegraphics[width = 4.25cm]{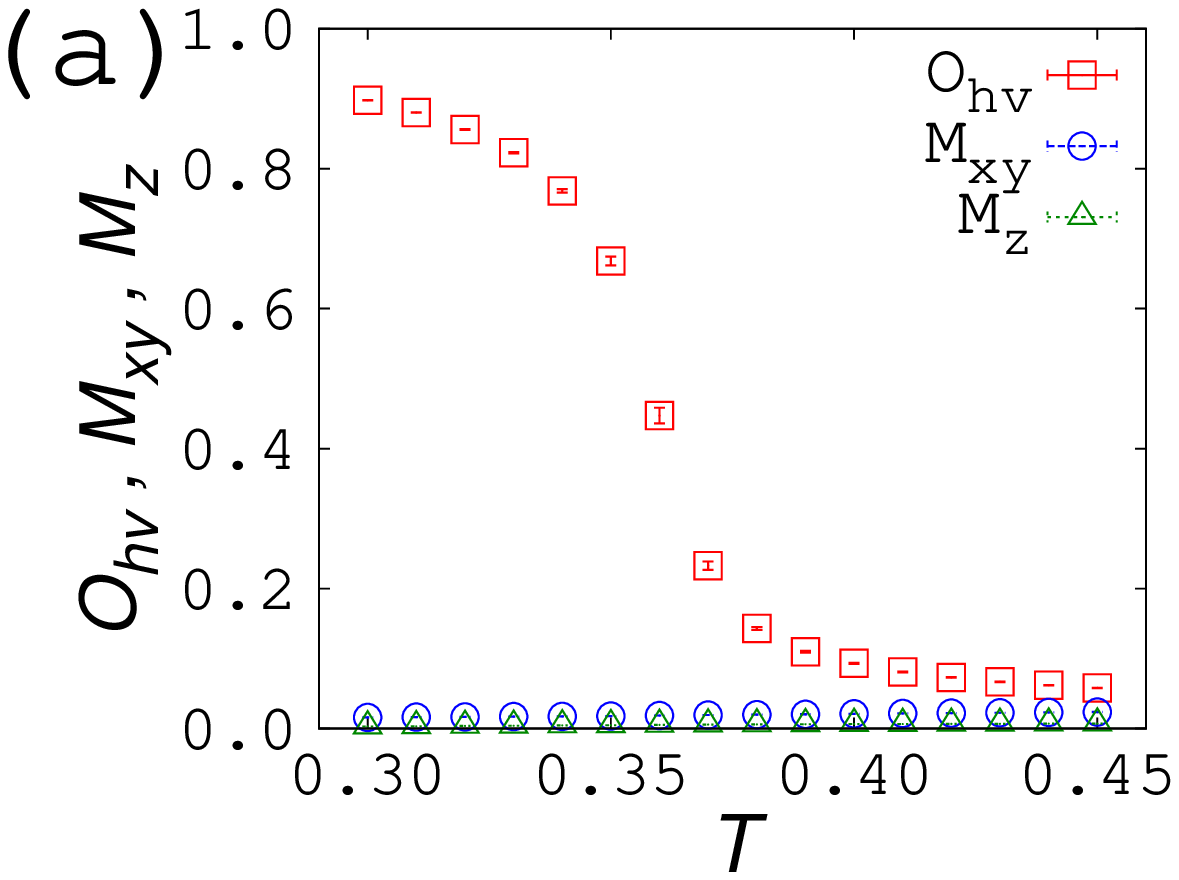}
\includegraphics[width = 4.25cm]{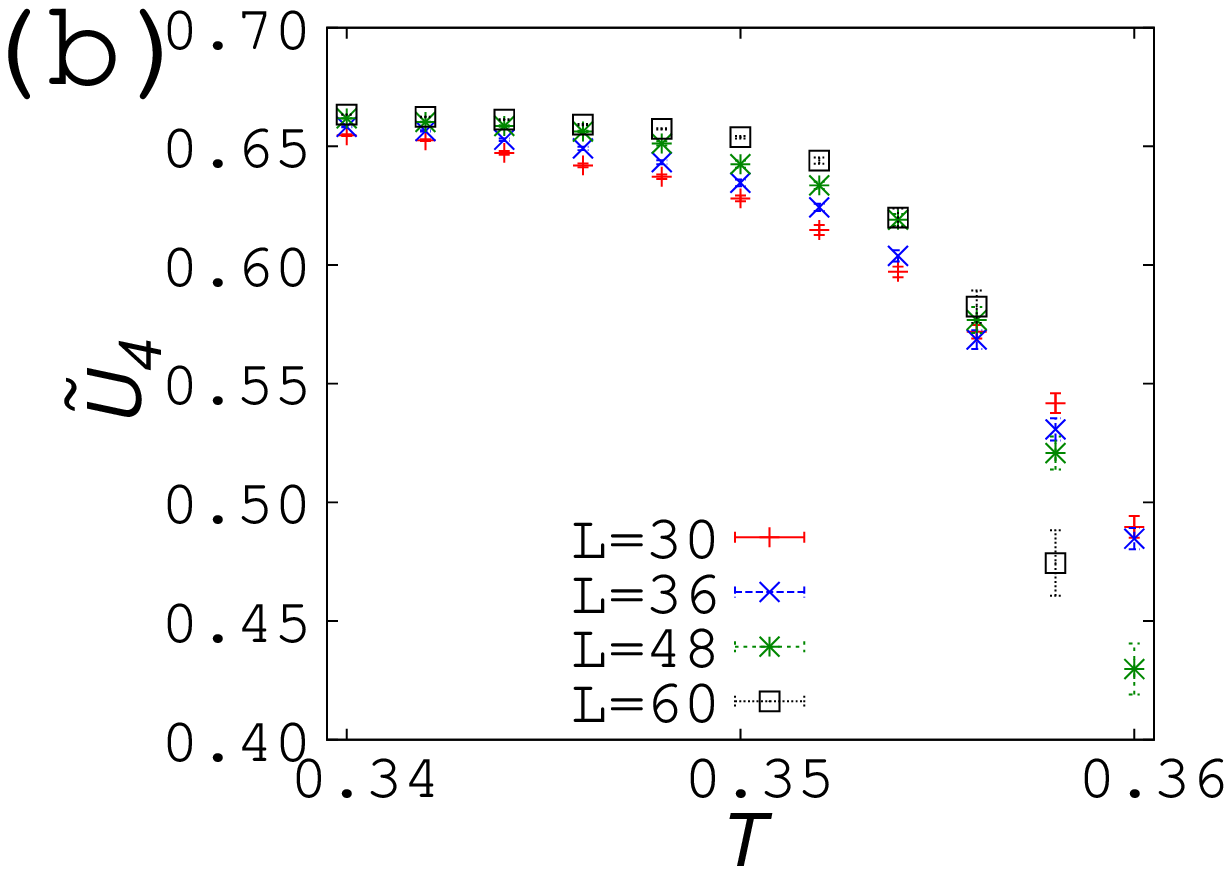}
\end{center}
\caption{Temperature dependence at $H=0$ and $\eta = 5$ of (a) $O_{hv}$ (red squares), $M_{xy}$ (blue circles), and $M_z$ (green triangles) for $L=30$, and (b) $\tilde{U_4}$ for $L=30$, $36$, $48$ and $60$.}
\label{T_vs_OD_U4_H0_eta5}
\end{figure}

\begin{figure}[tb]
\begin{center}
\includegraphics[width = 6.0cm]{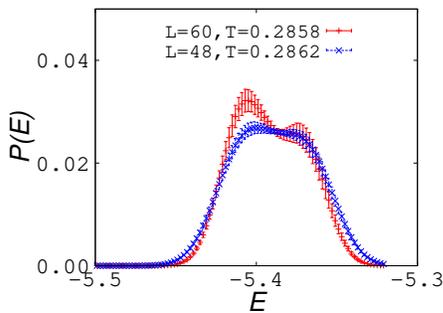}
\end{center}
\caption{Energy histogram $P(E)$ at $H=2$ and $\eta = 5$ near the transition point between the paramagnetic and $\left< 21 \right>$ stripe-ordered phases. 
The red and blue symbols denote $P(E)$ for $L=60$ at $T=0.2858$ and for $L=48$ at $T=0.2862$, respectively.}
\label{histo_H2_eta5}
\end{figure}

\subsection{Phase transition between the paramagnetic and stripe-ordered phases}
\label{Stripe-D}

In this subsection we study the phase transition between the P and SO phases, which is observed at relatively large $\eta$. 
For $H=0$, we find that the phase transition between the P and $\langle 1 \rangle$ SO phases (see Fig.~\ref{PD}(a)) is of second order. 
Figure~\ref{T_vs_OD_U4_H0_eta5}(a) illustrates the temperature dependence of $O_{hv}$, $M_{xy}$, and $M_z$ at $H=0$ and $\eta=5$, and Fig.~\ref{T_vs_OD_U4_H0_eta5}(b) shows the temperature dependence of the Binder cumulant $\tilde{U}_4$ for $O_{hv}$ for $L=30$, $36$, $48$ and $60$. 
 The order parameter $O_{hv}$ changes continuously and $\tilde{U}_4$ clearly crosses at $T \approx 0.356$, which indicates a second-order transition at $T_{\rm c}=0.356(2)$. 
In the same manner we determine the second-order transition points on the P--SO phase boundary in Fig.~\ref{PD}(a). This second-order transition for $\delta=1$ shows similar tendency to the result for $\delta=3$ by MacIsaac {\it et al}.~\cite{MacIsaac1} and not to that by Carubelli {\it et al}.~\cite{Carubelli}. 
 
We found the second-order transition in the Ising limit in our previous study~\cite{Komatsu}, and the second-order P--SO transition line is naturally extended to the Ising limit ($\eta \rightarrow \infty$). 
We notice that the critical temperature is little affected by the value of $\eta$, and that the values of the Binder cumulant at the crossing points, $\tilde{U}_4 \approx 0.55$, are the same as that in the Ising limit~\cite{Komatsu}.
For $\eta \gtrsim 4.4$, the model is considered as the Ising dipolar one. As we pointed out in Ref.~\cite{Komatsu}, there have been controversial results about the phase boundary for $\delta=1$ and $H=0$ in the Ising dipolar ferromagnet, i.e., first order~\cite{Cannas,Pighin-Ising,Rastelli} or second order~\cite{Fonseca,Horowitz,Bab}. 
The present analysis also supports ``second-order transition".

For finite fields ($H \ne 0$), the nature of the phase boundary depends on its value. For $H=1.3$, the SO phase has the $\langle 21^3 \rangle$ stipe pattern, and the boundary between the P and SO phases is of second order as well (see Fig.~\ref{PD}(b)). For $H=2$, however, the boundary between the P and $\langle 21 \rangle$ SO phases is of first order (see Fig.~\ref{PD}(c)), which is identified by the energy-histogram analysis. 
As an example, we show in Fig.~\ref{histo_H2_eta5} the energy histograms around the transition temperature ($T \approx 0.286$ with a slight size dependence) at $\eta=5$, in which double peaks exist and the peaks become sharper for larger $L$. The transition temperature hardly depends on the value of $\eta$ as well as in the $H=0$ case.

\subsection{Temperature-induced reentrant phase transition}
\label{Reentrant}

We find a temperature-induced reentrant phase transition: P phase $\rightarrow$ planar F phase $\rightarrow$ P phase. The reentrant transition is observable in the following four cases of successive transitions with lowering temperature. 

\noindent
Case I: P phase $\rightarrow$ planar F phase $\rightarrow$ P phase $\rightarrow$ SO $\langle 1 \rangle$ phase.

\noindent
Case II: P phase $\rightarrow$ planar F phase $\rightarrow$ P phase $\rightarrow$ SO $\langle 21^3 \rangle$ phase.

\noindent
Case III: P phase $\rightarrow$ planar F phase $\rightarrow$ P phase $\rightarrow$ SO $\langle 21 \rangle$ phase.

\noindent
Case IV: P phase $\rightarrow$ planar F phase $\rightarrow$ P phase. 

\begin{figure}[hbp!]
\begin{center}
\includegraphics[width = 4.25cm]{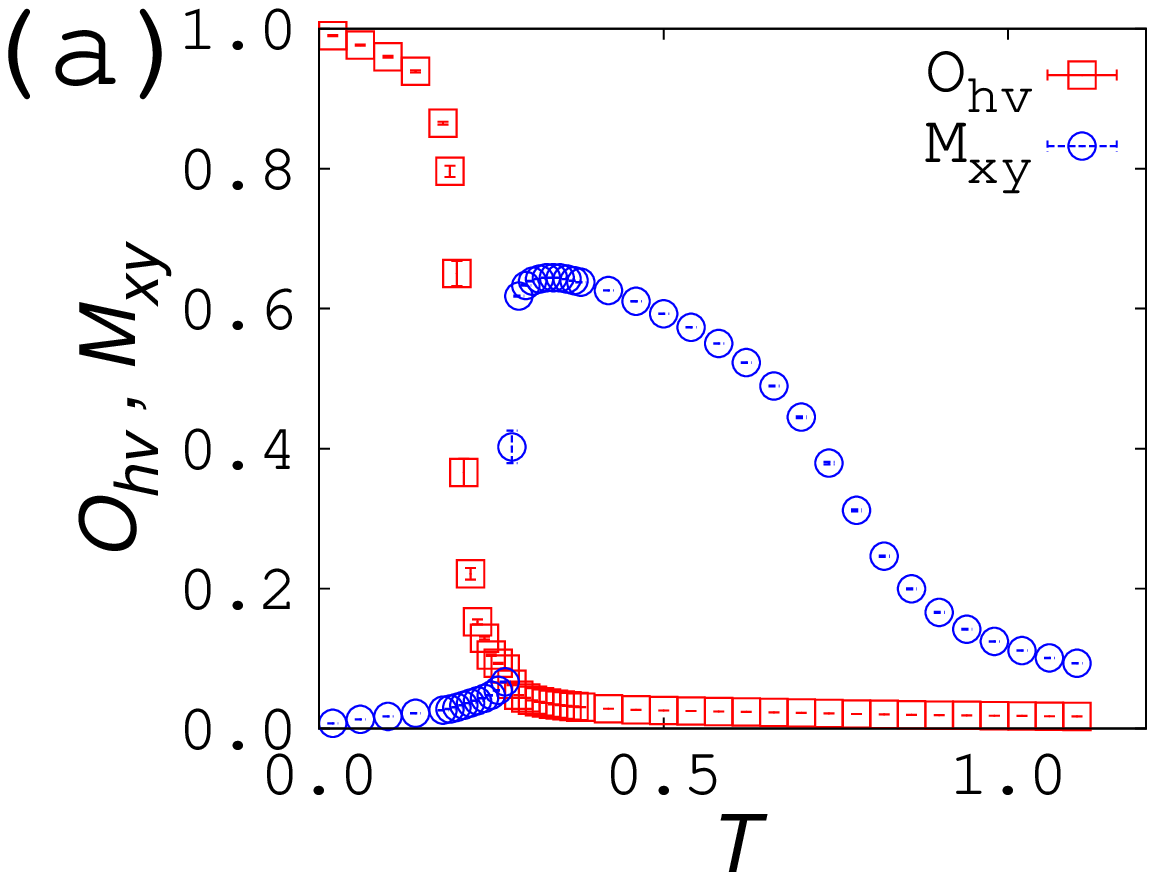}
\includegraphics[width = 4.25cm]{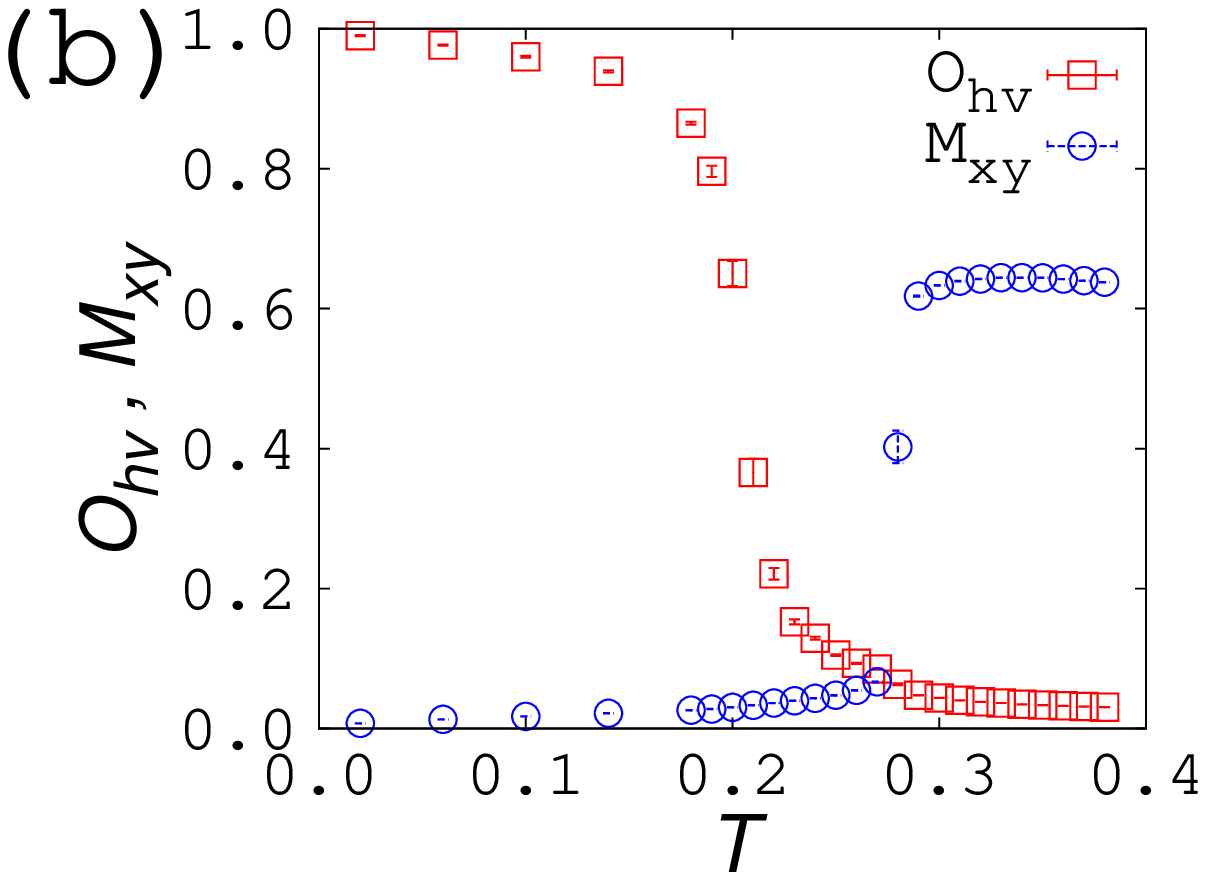}
\end{center}
\caption{Temperature dependence of $M_{xy}$ and $O_{hv}$ at $H=1.3$ and $\eta=4.2$ for $L=30$ in the (a) overall temperature region and (b) region $0<T<0.4$. The reentrant transition occurs as paramagnetic $\rightarrow$ planar ferromagnetic $\rightarrow$ paramagnetic $\rightarrow$ SO $\langle 21^3 \rangle$ phase with lowering temperature.}
\label{M_eta_4.2}
\end{figure}

\begin{figure}[hbp!]
\begin{center}
\includegraphics[width = 4.25cm]{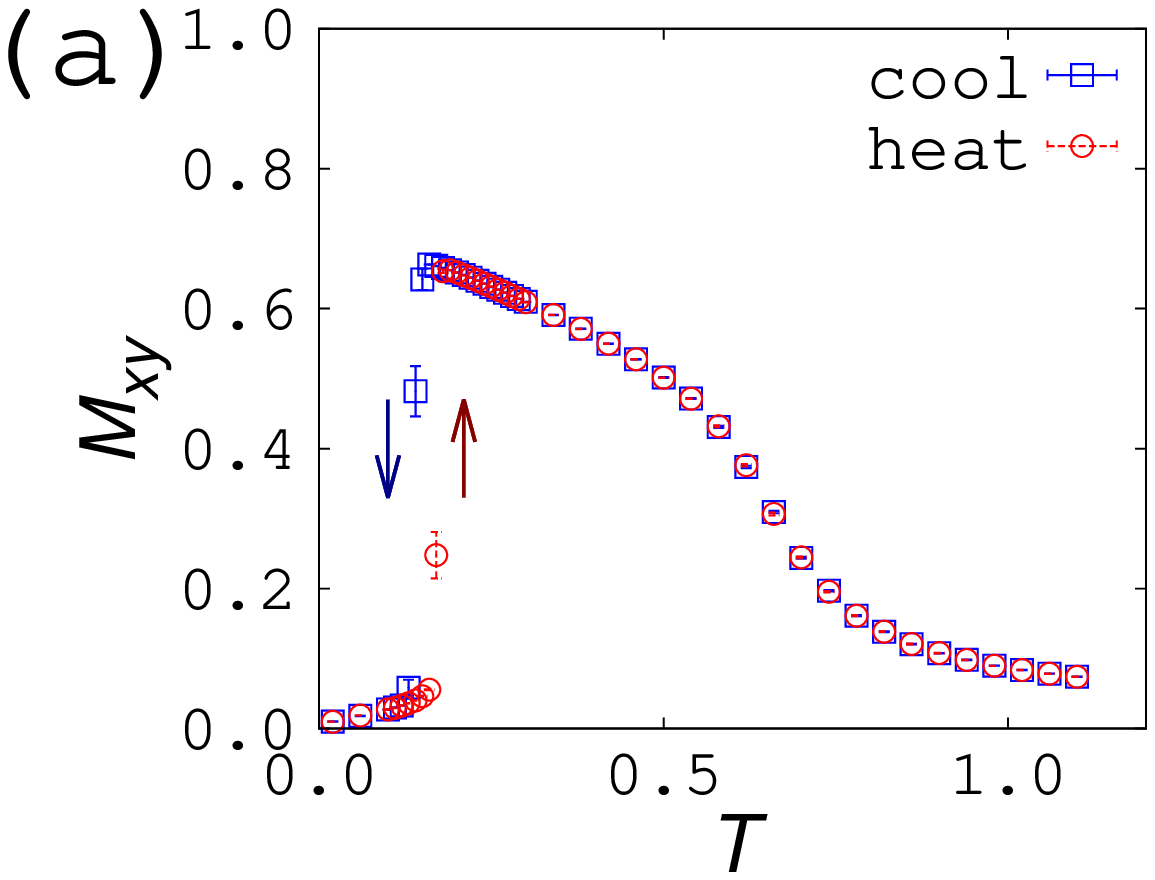}
\includegraphics[width = 4.25cm]{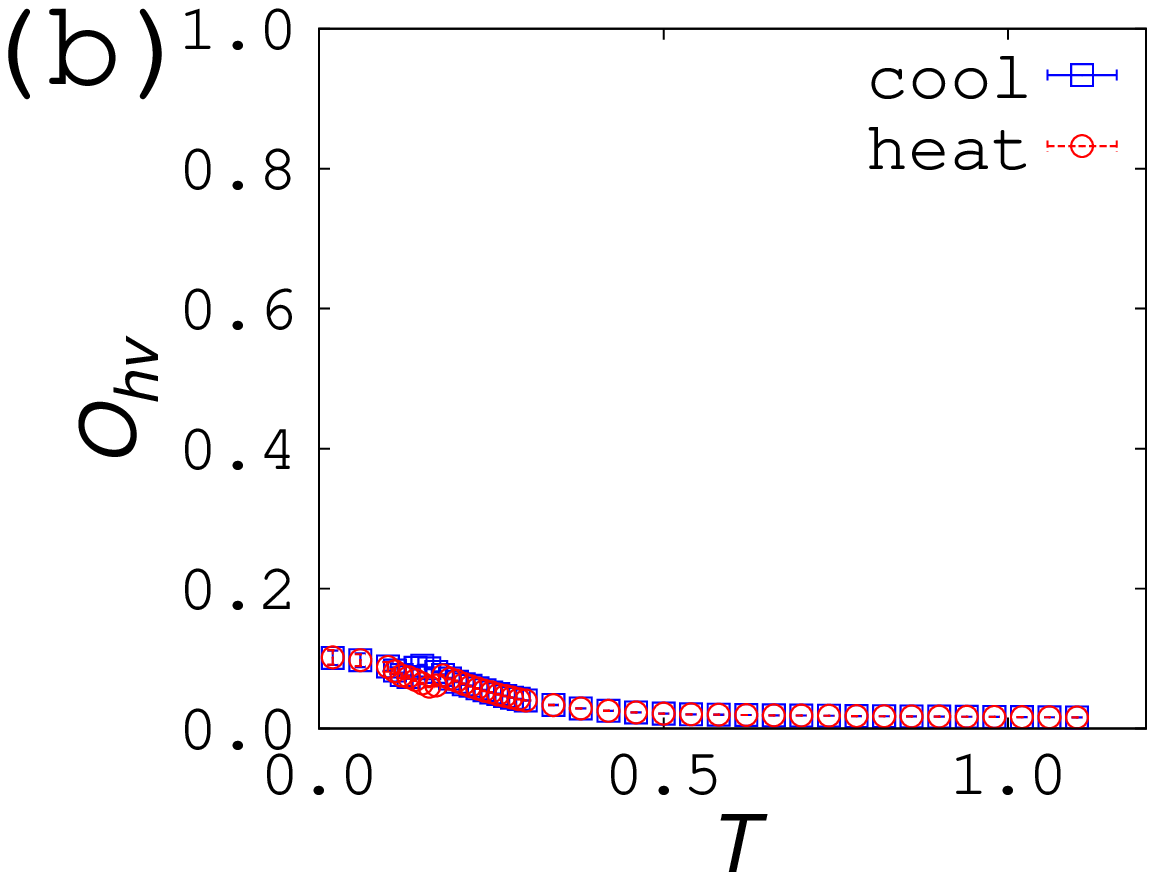}
\end{center}
\caption{Temperature dependence of (a) $M_{xy}$ and (b) $O_{hv}$ at $H=3.2$ and $\eta=4.1$ for $L=30$. The reentrant transition occurs as paramagnetic $\rightarrow$ planar ferromagnetic $\rightarrow$ paramagnetic phase with lowering temperature. A hysteresis loop is observed around the transition point between the paramagnetic and planar ferromagnetic phases.}
\label{M_eta_4.1}
\end{figure}

Cases I, II, III, and IV occur for $4.20 \lesssim \eta \lesssim 4.38$ at $H=0$, $4.10 \lesssim \eta \lesssim 4.30$ at $H=1.3$, $4.15 \lesssim \eta \lesssim 4.28$ at $H=2$, and $3.97 \lesssim \eta \lesssim 4.25$ at $H=3.2$, respectively. 
We give examples for cases II and IV  (cases I and III are not shown).
For case II, we show the temperature dependence of $M_{xy}$ and $O_{hv}$ at $\eta=4.2$ and $H=1.3$ in  Figs.~\ref{M_eta_4.2}(a) and \ref{M_eta_4.2}(b), respectively. 
With lowering temperature, $M_{xy}$ grows below $T \approx 0.80$ and disappears at $T \approx 0.28$, and $O_{hv}$ grows below $T \approx 0.22$. 
The P phase exists for $0.22 \lesssim T \lesssim 0.28$. 
For case IV, we depict the temperature dependence of $M_{xy}$ and $O_{hv}$ at $\eta=4.1$ and $H=3.2$ in Figs.~\ref{M_eta_4.1}(a) and \ref{M_eta_4.1}(b), respectively. The stripe order $O_{hv}$ virtually vanishes (small values at low temperatures  are due to a finite-size effect). 
With lowering temperature, $M_{xy}$ grows below $T \approx 0.70$, and disappears at $T \approx 0.10$ with hysteresis. The transition from the planar F to P phases at $\eta=4.1$ and $H=3.2$ is of first order, while it is of second order at $\eta=4.2$ and $H=3.2$. 
It is noting that in case IV the ground state is not an ordered state.

\section{Summary}
\label{summary}

We studied details of the $\eta$--$T$ phase diagrams for the 2D dipolar Heisenberg model with the uniaxial anisotropy $\eta$ with and without magnetic field for $\delta=1$ based on finite-size-scaling analyses. 
To obtain the phase diagrams, we used the SCO $O(N)$ MC algorithm for long-range interaction spin models with a modification efficient for models with the uniaxial anisotropy. 

At lower fields ($H < 2.8$), the phase diagram is characterized by the triple point at which the three phases encounter, i.e., the planar ferromagnetic (F), stripe-ordered (SO), and paramagnetic (P) phases. 
The stripe pattern of the SO phase varies with rising field as $\langle 1 \rangle \rightarrow \langle 21^3 \rangle \rightarrow  \langle 21  \rangle$ (Fig.~\ref{PD_magnified}). 
On the other hand, at higher fields ($H > 3.0$), any SO phase does not appear, while the planar F phase survives.

There are controversial results for $\delta=3$ at $H=0$ about the nature of the transition between the P and $\langle 4 \rangle$ SO phases~\cite{MacIsaac2,Carubelli}. 
Concerning $\delta=1$, we found that the transition between the P and $\langle 1 \rangle$ SO phases is of second order at $H=0$, and the order of the transition varies depending on the value of the field, i.e., it is of second order between the P and $\langle 21^3 \rangle$ SO phases at $H=1.3$ but of first order between the P and $\langle 21 \rangle$ SO phases at $H=2$.

We investigated the spin-reorientation (SR) transition between the planer F and SO phases and confirmed that this transition is of first order, which agrees with previous studies for different $\delta$~\cite{Pescia,Hucht,MacIsaac1,MacIsaac2,Carubelli,Santamaria}. 
We found that the slope of the SR transition line in the $\eta$--$T$ diagram 
is positive at zero field, i.e., $\frac{d\eta}{dT}>0$. 
There exists controversy on the sign of $\frac{d\eta}{dT}$ for $\delta=0$~\cite{MacIsaac1,Carubelli} and $\delta=3$~\cite{MacIsaac2,Carubelli}. Considering the present result for $\delta=1$, previous studies for $\delta \gg 1$~\cite{Moschel,Pescia}, and experimentally-observed transitions from the in-plane ferromagnetic phase to the out-of-plane stripe-ordered phase with decreasing temperature~\cite{Pappas,Allenspach,Ramchal,Won,Qiu}, $\frac{d\eta}{dT}>0$ for the SR transition would be more plausible. 

In all cases with and without field, we found a characteristic shape of the transition line between the P and planar F phases, whose maximum point of $\eta$ is located at an intermediate temperature in the $\eta$--$T$ phase diagram. The P--planer F phase transition line is of second order in a wide region of high temperatures. We found a tricritical point on the transition line at a low temperature at $H=1.3$ and $H=3.2$. However, we could not find it at $H=0$ and $H=2$ within the accuracy of the present simulation. 

The characteristic shape of the P--planer F transition line causes a temperature-induced reentrant transition: P phase to planar F phase to P phase. So far reentrant transitions have been studied in systems with competing short-range interactions~\cite{Kitatani} as observed in spin glasses~\cite{Maletta,Aeppli,Childress}. 
Here we found a novel reentrant transition due to the competition between the uniaxial anisotropy, short-range and long-range interactions. This reentrant transition is observable in the following four cases of successive transitions with lowering temperature: P phase $\rightarrow$ planar F phase $\rightarrow$ P phase $\rightarrow$ SO phase ($\langle 1 \rangle$, $\langle 21^3 \rangle$, or  $\langle 21  \rangle$) at lower fields, and P phase $\rightarrow$ planar F phase $\rightarrow$ P phase at higher fields. 
A similar shape of the P--planer F phase transition line was reported for $\delta=3$ and $H=0$~\cite{Carubelli}, and the reentrant transition might be observed in that case: P phase $\rightarrow$ planar F phase $\rightarrow$ P phase $\rightarrow$ SO $\langle 4 \rangle$ phase. 
We hope that the present study promotes further researches for fully understanding of the ultrathin film magnetism.

\section*{Acknowledgments}
The present study was supported by Grants-in-Aid for Scientific Research C (No. 17K05508) from MEXT of Japan, and the Elements Strategy Initiative Center for Magnetic Materials (ESICMM) under the outsourcing project of MEXT. Part of numerical calculations were performed on the Numerical Materials Simulator at National Institute for Materials Science.

\end{document}